\documentclass{article}

\usepackage{amsthm}
\usepackage{amsmath}
\usepackage{amsfonts}
\usepackage{amssymb}
\usepackage{braket}
\usepackage{complexity}
\usepackage[margin=1in, marginpar=0.7in]{geometry}
\usepackage{graphicx}
\usepackage[pagebackref=true]{hyperref}
\usepackage{mathtools}
\usepackage[numbers,sort&compress]{natbib}
\usepackage{thm-restate}

\makeatletter
\newcommand{\turnoffnumbering}{
\global\st@rredtrue
\global\@eqnswfalse
}
\newcommand{\turnonnumbering}{
\global\st@rredfalse
\global\@eqnswtrue
}
\makeatother

\newcommand{\maintitle}{Electronic Structure in a Fixed Basis is QMA-complete}
\newcommand{\authors}{
\author{
Bryan O'Gorman\thanks{UC Berkeley \& NASA Ames,  ~$\dag$UC Irvine, ~$^{**}$Dartmouth College, ~$\ddag$University of Chicago}
\and
Sandy Irani$^\dag$
\and
James Whitfield$^{**}$
\and
Bill Fefferman$^{\ddag}$
}
}

\newcommand{\ham}[2][H]{#1^{(\mathrm{#2})}}
\newcommand{\degree}{\mathrm{deg}}
\newcommand{\orbitalcenters}{X}

\newcommand{\norm}[1]{\left\lVert#1\right\rVert}

\newcommand{\block}{\mathcal B}
\newcommand{\on}{\mbox{\sc{On}}}
\newcommand{\off}{\mbox{\sc{Off}}}
\DeclareMathOperator\erf{erf}

\newtheorem{definition}{Definition}
\newtheorem{theorem}{Theorem}
\newtheorem*{result}{Theorem}
\newtheorem{corollary}{Corollary}
\newtheorem{lemma}{Lemma}

\hypersetup{
    colorlinks=true,
    pdftitle={\maintitle}
}

\usepackage[capitalise]{cleveref}

\newtheorem{innercustomthm}{Theorem}
\newenvironment{customthm}[1]
  {\renewcommand\theinnercustomthm{#1}\innercustomthm}
  {\endinnercustomthm}

\authors

\date{\today}
\title{
\maintitle
}

\begin{document}
\maketitle

\begin{abstract}
Finding the ground state energy of electrons subject to an external electric field is a fundamental problem in computational chemistry.
We prove that this electronic-structure problem, when restricted to a fixed single-particle basis and fixed number of electrons, is QMA-complete.
Schuch and Verstraete have shown hardness for the electronic-structure problem 
 with an additional site-specific external \emph{magnetic} field, but without the restriction to a fixed basis
\cite{schuch2009computational}.
In their reduction, a local Hamiltonian on qubits is encoded in the site-specific magnetic field.
In our reduction, the local Hamiltonian is encoded in the choice of
spatial orbitals used to discretize the electronic-structure Hamiltonian.
As a step in their proof, Schuch and Verstraete show a reduction from the antiferromagnetic Heisenberg Hamiltonian
to the Fermi-Hubbard Hamiltonian. We combine this reduction with the fact that the antiferromagnetic Heisenberg Hamiltonian 
is QMA-hard \cite{piddock2017complexity} to observe that the Fermi-Hubbard Hamiltonian on generic graphs is QMA-hard, even when all the hopping coefficients have the same sign.
We then reduce from Fermi-Hubbard by showing that an instance of Fermi-Hubbard can
be closely approximated by an instance of the Electronic-Structure Hamiltonian in a fixed basis.
Finally, we show that estimating the energy of the lowest-energy Slater-determinant state
(i.e., the Hartree-Fock state) is NP-complete for the Electronic-Structure Hamiltonian in a fixed basis.
\end{abstract}

\section{Introduction}

Simulating quantum mechanical systems is one of the most important computational challenges in modern science.  
Solving this problem, broadly defined, will allow us to probe the foundations of physics, chemistry, and materials science, and will have useful applications to a wide variety of industries. 
On the other hand, the very properties that make quantum mechanical systems so interesting -- such as the exponential growth of the underlying state space and quantum entanglement -- also make quantum simulation a particularly difficult computational task. 

Finding the means to tame this daunting complexity is an objective that is nearly as old as quantum mechanics itself.  
Paul Dirac, in a foundational paper from 1929, asserted that ``The fundamental laws necessary for the mathematical treatment of a large part of physics and the whole of chemistry are thus completely known, and the difficulty lies only in the fact that application of these laws leads to equations that are too complex to be solved. 
It therefore becomes desirable that approximate practical methods of applying quantum mechanics should be developed, which can lead to an explanation of the main features of complex atomic systems without too much computation'' \cite{Dirac29}.  

Today, nearly a century later, Dirac's quote captures the underlying motivation for a large body of quantum science research. 
For example, in the context of simulating systems of many electrons, the complexity inherent in the simulation problem has been addressed by approximation methods such as Hartree-Fock and Density Functional Theory~\cite{helgaker2000molecular}, as well as by considering simplified quantum models such as the Hubbard and Heisenberg Hamiltonians \cite{Ashcroft}.  Moreover, even in a new, exciting era in which noisy, intermediate-scale quantum computers are being developed that may be well suited to solve certain quantum simulation problems, Dirac's wisdom prevails.  Existing quantum algorithms, such as the phase estimation algorithm and the variational quantum eigensolver all obtain approximate solutions to special cases of the quantum simulation problem (see, e.g., \cite{Preskill2018,Kitaev96,Peruzzo2014}).

However, there are also fundamental limitations to these simulation algorithms that stem from quantum computational complexity.  Kitaev, building on the classical work of Cook and Levin, proved that a very general quantum simulation problem (approximating the ground state energy of a $k$-local Hamiltonian) is QMA-complete \cite{kitaev02,Cook71}.  QMA is a natural quantum analogue of NP, and QMA-complete problems should not have an efficient quantum algorithm, for essentially the same reasons that NP-complete problems (such as boolean satisfiability) should not have efficient classical algorithms. 

Nonetheless, QMA-completeness should not be interpreted as a categorical roadblock, but rather as an important guidepost for the development of future quantum algorithms.  In the same way that many practically interesting instances of classical constraint-satisfaction problems have special structural properties that avoid the worst-case hardness implied by NP-completeness results, we study QMA-completeness in order to understand which structural properties reduce the complexity of the simulation problem -- and which properties do not -- enabling improved quantum simulation algorithms that could potentially exploit this structure. 

In this work our goal is understand the computational difficulty of simulating systems of interacting elections.  Our main result shows that when restricted to a fixed number of electrons and a fixed single-particle basis, approximating the ground state energy of the electronic structure Hamiltonian is QMA-complete.  This can be interpreted as a direct sharpening of Dirac's quote: we conclusively demonstrate that these properties \emph{do not} add enough structure to enable the existence of an efficient quantum simulation algorithm to approximate the ground state energy of such systems.

\subsection{Paper Outline} 

Section \ref{sec:results} gives an overview of our results and techniques.
Section \ref{sec:fermihubb} gives the reduction from the antiferromagnetic
Heisenberg Hamiltonian to the Fermi-Hubbard Hamiltonian.
Section \ref{sec:ES} shows the reduction from Fermi-Hubbard to the
Electronic Structure Hamiltonian in a fixed basis.
In Section \ref{sec:hf} we show that finding the lowest energy
Hartree-Fock state for the Electronic Structure Hamiltonian in a fixed basis
is NP-complete.
Some of the technical lemmas used to show the QMA-hardness of
the Electronic Structure Hamiltonian in a fixed basis are given in the appendix.

\section{Overview of Results}
\label{sec:results}
\subsection{Formalizing the electronic structure problem}
In computational complexity, finding the ground state of a local Hamiltonian acting on qudits is the canonical QMA-complete problem. 
Local Hamiltonians are interesting not only because they are analogous to classical constraint-satisfaction problems involving a set
of low-arity functions, but because physical Hamiltonians are typically local due to the nature of physical
forces. 
The computational complexity of finding ground states of qubit Hamiltonians has been studied extensively, and hardness shown even for Hamiltonians that are physically realistic in the sense, e.g., that the terms are placed on a 2D lattice or all the same up to a positive rescaling~\cite{piddock2017complexity}.
Much less is known about the hardness of local Hamiltonians for indistinguishable particles. 
In this work, we consider the local Hamiltonian problem for fermionic systems.

The local Hamiltonian problem for systems of indistinguishable particles has two distinctive features.
First, the Hamiltonians themselves are invariant under permutations of the particles.
Second, the goal is to estimate the lowest-energy of a symmetric (for bosons) or anti-symmetric  (for fermions) state.
Generic Hamiltonians (i.e., quartic polynomials in the elementary operators with general coefficients) for both types of indistinguishable particles have been shown to be QMA-complete~\cite{wei2010interacting,liu2007quantum}, but, as with Hamiltonians on distinguishable particles, we can ask how hard more physically realistic classes of Hamiltonians are.
Physically realistic Hamiltonians on indistinguishable particles have special properties that could make them more amenable to computing ground energies.
In particular, here we are focused on the computational complexity of the \emph{electronic structure Hamiltonian}
\begin{equation}
\ham{ES} = 
-\frac12 \sum_i \nabla^2_i 
+\sum_i V(\mathbf r_i)
+
\frac12 \sum_{i \neq j} \frac{1}{\left|\mathbf r_i- \mathbf r_j\right|},
\end{equation}
which acts on an anti-symmetric state $\psi \in {\mathbb R}^{\eta \times 3}$ of $\eta$ electrons, where $\mathbf r_i$ is the position of the $i$-th electron in $3$-dimensional space.
For $\eta$ electrons and a specified electric potential $V: {\mathbb R}^3 \to \mathbb R$, this is the Hamiltonian dictated by the laws of electromagnetism.
Of particular interest in chemistry is the \emph{molecular electronic structure Hamiltonian}, in which the external potential
\begin{equation}
V(\mathbf r) = -\sum_{j} \frac{Z_j}{\left|\mathbf r - \mathbf R_j\right|}
\end{equation}
is that of nuclei modelled as classical point particles, each with positive charge $Z_j$ and located at fixed position $\mathbf R_j$.
In reality, the nuclei are also quantum particles, but they are so much more massive than the electrons that this model (the \emph{Born-Oppenheimer} approximation)
is usually a sufficiently accurate approximation  to the Hamiltonian of a molecule specified by the nuclear charges and number of electrons. 
There is a separate optimization procedure to find the lowest-energy configuration of nuclear positions.

Physically, the wavefunction of the electrons is over continuous real space.
Computationally, we need to  discretize the space of
possible wavefunctions in some way in order to have a finite representation
of a potential ground state.
This leads to the fundamental computational problem of quantum chemistry, estimating the ground state energy of the \emph{electronic structure Hamiltonian in a fixed basis}:
\begin{equation}\label{eq:es}
\ham{ES}(\boldsymbol \phi, V)
=
T + V + U
=
\sum_{\substack{i, j \in [n] \\ \sigma \in \{\pm 1\}}}
t_{i, j} a_{i, \sigma}^{\dagger} a_{j, \sigma}
+
\sum_{\substack{i, j \in [n] \\ \sigma \in \{\pm 1\}}}
v_{i, j} a_{i, \sigma}^{\dagger} a_{j, \sigma}
+
\frac12
\sum_{
    \mathclap{
        \substack{i, j, k, l \in {[n]} \\ \sigma, \tau \in {\{\pm 1\}}}}}
u_{i, j, k, l}
a_{i, \tau}^{\dagger}
a_{j, \sigma}^{\dagger}
a_{k, \sigma}
a_{l, \tau}
\end{equation}
where $\boldsymbol \phi = (\phi_1, \ldots, \phi_n)$ is the single-particle basis with elements $\phi_i: {\mathbb R}^3 \to \mathbb C$ and
\begin{align}
    t_{i, j} &= -\frac12 \int d \mathbf r \phi^*_i(\mathbf r) \nabla^2 \phi_j(\mathbf r), \label{eqn:t}\\
    v_{i, j} &= \int d \mathbf r \phi^*_i(\mathbf r) V(\mathbf r) \phi_j(\mathbf r), \label{eqn:v}\\
u_{i, j, k, l}
    &=
\int d \mathbf r d \mathbf s
\phi^*_i(\mathbf r) \phi^*_j(\mathbf s)
\frac{1}{\left|\mathbf r - \mathbf s\right|}
\phi_k(\mathbf s) \phi_l(\mathbf r).
\label{eqn:u}
\end{align}
The indices $i, j, k, l \in [n]$ index spatial orbitals, and $\sigma, \tau \in \{\pm 1\}$ indicate the spin.
Given the potential $V(\mathbf r)$ and a fixed set of orbitals, the Hamiltonian shown in (\ref{eq:es}) is then completely determined by the
integrals for the kinetic and potential energy shown in \cref{eqn:t,eqn:v,eqn:u}.
The {\em Electronic Structure} problem  then is to determine whether the ground energy of the resulting Hamiltonian is less than some threshold $E$
or greater than $E+1/\poly(n)$.
This is the version of the problem posed by Whitfield et al. (\cite{whitfield2013computational}), who left its hardness as an open problem.
We answer here in the affirmative by showing a family of single-particle bases (with zero potential $V(\mathbf r) = 0$) that encodes hard problems.
In $\ham[H]{ES}$, the $\sigma$ and $\tau$ indicate the sign of the spin of the spin orbital.
For each spatial orbital $\phi_i(\mathbf r)$, there are two spin orbitals $\phi_{i, \pm 1}(\mathbf r)$.
We use $\pm1$ as an index for simplicity, but of course physically the electron's spin has magnitude $1/2$.

\begin{definition}[Electronic structure in fixed basis set -- ESFBS]\label{def:elec-struct}
An instance of electronic structure in a fixed basis set is specified by an external electric field $V: \mathbb R^3 \to \mathbb R$, a number $\eta$ of electrons, a basis set $\boldsymbol \phi = (\mathbf \phi_1, \ldots, \mathbf \phi_n)$, and thresholds $a < b$, where $b-a \ge 1 / \poly(\eta)$.
The external potential $V$ and the basis set $\boldsymbol \phi$ must be specified concisely (using $\poly(n)$ bits) in a way that allows for efficient ($\poly(n)$-time) calculation of the integrals in \cref{eqn:t,eqn:v,eqn:u}.
The goal is to determine whether the ground state energy of $H_{\mathrm{ES}}$ in the subspace of $\eta$ electrons spanned by the given basis is at most $a$ or at least $b$.
\end{definition}

Our definition of the problem allows for states with arbitrary total spin, and this freedom will be critical in our construction.
One can also consider a variant in which the total spin is fixed, analogous to, for example, the XY problem with fixed magnetization.
Our definition of the problem also restricts the states allowed to a finite-dimensional space spanned by a set of fixed single-electron orbitals.
By necessity, this is the form of the problem solved in practice by computational chemists.
However, for practical purposes, it is desirable that the ground state or ground energy in the chosen basis is close to that in a complete, infinite-dimensional basis.
The difference between these two is known as the basis-set error, and bases are typically chosen in order to minimize this error.
The basis we use in our construction is artificial in this sense; in the absence of an external potential, there is nothing to to confine the electrons to the subspace of $\mathbb{R}^3$ spanned by the basis. 
However, the orbitals that we use are still superpositions of Gaussians, a commonly used form in computational chemistry, e.g. the STO-3G basis set~\cite{hehre69} with each basis function composed of a fixed superposition of three primitive Gaussians. 
Indeed, we prove the following theorem 
that the electronic structure in a fixed
basis is QMA-hard by encoding a  QMA-hard Hamiltonian  in the construction of the basis.

\begin{theorem}[ESFBS is QMA-complete, informal]\label{thm:es-informal}
The electronic structure problem in a fixed-basis set and at fixed particle number is QMA-complete.
\end{theorem}

Our results contribute to a large body of work formally establishing the computational intractability of increasingly physically realistic Hamiltonians.
There are still many important problems in computational chemistry whose computational complexity is unknown.
For example, even in a fixed basis, does fixing the spin make the problem easier? Does the problem become more tractable if the given orbitals
are guaranteed to have small basis-set error?
Is the electronic structure problem hard in a complete (infinite-dimensional) basis?
If so, is it still hard when the external potential arises solely from a set of positively charged nuclei at fixed positions?
We pose two variants of the electronic structure problem whose hardness is an open question.
In both cases, the ``size'' of the problem is the number of electrons.
\begin{description}
\item 
{\bf Electronic structure in a fixed basis with bounded basis-set error:}
Given an external electric potential $V$, number of electrons $\eta$, thresholds $a < b$, and a basis set $\boldsymbol \phi$ with basis-set error $\epsilon(\eta) = 1 / \poly(\eta)$ for the given potential $V$, determine whether the lowest energy of a state in the space spanned by $\boldsymbol \phi$ is at most $a$ or greater than $b = a + 1 / \poly(\eta)$.
The basis-set error is defined as $\braket{\tilde{\psi} | H | \tilde{\psi}} - \min_{\ket{\psi} \in \mathbb R^{\eta \times 3}}\braket{{\psi} | H | {\psi}} \leq \epsilon(\eta)$.
\end{description}

The parameters of the \emph{problem} are the promised basis-set error bound $\epsilon(\eta)$, the thresholds $a(\eta)$ and $b(\eta)$, and the family of potentials considered (as a function of $\eta$); an instance is specified by simply the number of electrons $\eta$, the potential $V$, and a specification of the
basis set $\boldsymbol \phi$.
This variant entertains the possibility that, while the problem is hard for
arbitrary \emph{bases}, it may always be easy for \emph{good} bases (in the sense of having low basis-set error).
In practice, chemists always want to use a good basis, and often do, though in general they have no guarantees on the error of the bases that they use.
Note that a good basis need not necessarily be complete for the whole space; all that matters is that its span includes a state sufficiently close to the ground state.
For example, in Schuch and Verstraete's construction for the QMA-hardness of electronic structure with magnetic fields, the external electric potential $V$ implies a good basis of size $n = \eta$ that captures the ground state but is far from complete.
Theoretical and numerical results suggest that for physically realistic external potentials there is a always good basis of size $\poly(\eta)$~\cite{kato1957eigenfunctions,halkier1998basis}, though the constant prefactors may be impractically large.
Furthermore, there may exist pathological external potentials for which no polynomially large good basis exists.

To account for both the possibility of no good polynomially large basis and the desirability of working in a small basis, we define another variant of the problem that includes finding the basis in which the state is expressed.
The formulation attempts to be as general as possible while remaining in QMA. 
Ideally, we would like to consider all states that can be efficiently represented and whose energy can be efficiently estimated by a quantum computer.
To formalize this, we specify some family of parameterized orbitals in which the putative low-energy state can be expressed.
For example, the family of bases could consist of all weighted sums of Gaussians. 
In this case the prover would provide, for each basis element, the centers, the weights, and the exponents of the constituent Gaussians. 

\begin{description}
\item {\bf Electronic structure in parameterized basis:}
Given an external electric potential $V$, number of electrons $\eta$, thresholds $a < b$, and a family of basis functions $\left\{\phi_{\theta}\right\}_{\theta}$, and basis size $k$, determine whether there exists a basis $\boldsymbol \phi = (\phi_{\theta_1}, \ldots, \phi_{\theta_k})$ such that the lowest energy of a state in the space spanned by $\boldsymbol \phi$ is at most $a$ or greater than $b = a + 1 / \poly(\eta)$.
\end{description}
The problem is parameterized by the thresholds $a(\eta)$ and $b(\eta)$ and the family of basis functions ${\{\phi_{\theta}\}}_{\theta}$ allowed; an instance is specified by just the number of electrons $\eta$, basis set size $k$, and potential $V$.
A certificate consists of the classical description of orbitals $\boldsymbol \phi$ and a quantum state on $2k$ qubits that is supposed to represent a low-energy of state of $\eta$ electrons in the basis $\boldsymbol \phi$.

Other variants of the electronic structure problem have been considered.
Schuch and Verstraete show QMA-hardness for electronic structure with an additional
site-specific magnetic field \emph{in a good basis},  which is used to encode an instance of a QMA-hard problem~\cite{schuch2009computational}.
Their result is thus incomparable to ours; we removed the magnetic field, but also the restriction to a good basis.

There are several related computational problems concerning various ways of representing and working with quantum states.
One is $N$-representability.
Note that the 2-electron reduced density matrices (2-RDMs) of the quantum state encode all the information necessary to compute the energy of the electronic structure.
The $N$-representability problem is to determine whether or not for a given set of 2-RDMs there exists a consistent quantum state on the full space.
This problem has been shown, under Turing reductions, to be QMA-complete~\cite{liu2007quantum}.
Another related problem is Density Functional Theory (DFT), which is premised on the fact that the electron density (i.e. the average number of electrons at each point in space) is also sufficient to calculate the energy of the electronic structure Hamiltonian.
That is, there exists a universal functional that takes as input the electron density and outputs the energy.
However, while such a functional exists, it may not be computationally efficient.
Indeed, computing it has been shown to be QMA-hard, also under Turing reductions~\cite{whitfield2014computational}.
For both $N$-representability and DFT, it remains an open question whether they remain hard when the inputs (2-RDMs and electron densities, respectively) are restricted to the ground states of electronic structure Hamiltonians.
Broadbent and Grilo recently proved~\cite{broadbent2020qma} QMA-completeness of the Consistency of Local Density Matrices problem (i.e., the qudit analog of $N$-representability) under Karp reductions, but left as an open question whether or not their techniques can be used to show QMA-hardness under Karp reductions of $N$-representability and the universal functional of DFT.

\subsection{Hubbard Hamiltonians}

The proof that   Electronic Structure in a Fixed Basis Set  (ESFBS)
is QMA-hard proceeds in two stages. We first reduce from 
the antiferromagnetic Heisenberg Hamiltonian to the Fermi-Hubbard
Hamiltonian. Then we reduce from Fermi-Hubbard to ESFBS.
This subsection gives an overview of the first reduction.

The Bose-Hubbard and Fermi-Hubbard Hamiltonians are:
\begin{align}
\ham{BH}
&=
\sum_{i \in V} U n_i (n_i - 1) 
+
\sum_{\mathclap{\{i, j\} \in E}} t_{i, j} \left(b_{i}^{\dagger} b_{j} + \text{h.c.}\right),
&
\ham{FH}
&=
\sum_{i \in V} U n_{i, +} n_{i, -}
+
\sum_{\mathclap{\{i, j\} \in E, \sigma \in \{\pm\}}}
t_{i, j}
a_{i, \sigma}^{\dagger} a_{j, \sigma}
,
\end{align}
where $G = (V, E)$ is the interaction graph, and $a_ia_j^\dag + a_j^\dag a_i= b_ib_j^\dag - b_j^\dag b_i =\delta_{ij}$.
When we refer to the ``Hubbard'' model without qualification, we mean the \emph{Fermi}-Hubbard model, in which the particles are fermions.
Hubbard Hamiltonians are of practical interest because they approximate  Hamiltonians of many more complicated condensed-matter and chemical systems.
Their solutions are taken to qualitatively describe those of the approximated systems.

Childs et al.~\cite{childs2015bose, childs2016complexity} show that the Bose-Hubbard Hamiltonian and XY Hamiltonian are QMA-hard with uniform coefficients.
In both cases, because the coefficients are uniform, the instance is encoded entirely in the graph, which does not seem embeddable in, say, three spatial dimensions, as we would want for a physically realistic Hubbard Hamiltonian.
Schuch and Verstraete~\cite{schuch2009computational} show as an intermediate result that the Fermi-Hubbard Hamiltonian on a 2D lattice with a site-specific \emph{magnetic field} is QMA-hard; the instance is encoded entirely in this magnetic field.
We show that the magnetic field is not necessary, at the cost of having
an arbitrary weighted interaction graph.

\begin{result}[FH is QMA-complete]\label{res:hubbard}
The Fermi-Hubbard Hamiltonian with arbitrary coefficients and  fixed particle number is QMA-complete,
even if all of the tunneling coefficients have the same sign and are bounded
by a polynomial in the number of particles.
\end{result}

The proof reduces from the antiferromagnetic Heisenberg Hamiltonian:
\begin{align}
\ham{Heis} &= \sum_{\{i, j\} \in E} \kappa_{i, j} W_{i, j}
,
&
W 
&=
(II + XX + YY + ZZ) / 2
,
\end{align}
which is known to be QMA-hard \cite{cubitt2016complexity, piddock2017complexity}. 
As in related previous constructions, we fix the number of particles to equal the number of spatial orbitals, i.e., half the number of spin orbitals.
The large onsite-repulsion term $U$ penalizes two electrons occupying the same spatial orbital, and so the ground space of the repulsion term has exactly one electron in each spatial orbital.
As was done in \cite{liu2007quantum}, the spin of the electron in each orbital encodes a logical qubit.
With the repulsion term dominating the Hamiltonian, we treat the rest perturbatively.
To second order, this yields an antiferromagnetic Heisenberg Hamiltonian on the same graph as a Hubbard Hamiltonian.
We go between a qubit Hamiltonian and a fermionic Hamiltonian
using the Jordan-Wigner transformation 
$a_i \leftrightarrow \prod_{j<i} Z_j (X_i + i Y_i) /2$.
In general, this transforms local fermionic Hamiltonians into non-local qubit Hamiltonians, but with a particular ordering of the spin orbitals, the parity strings $\prod_{j<i} Z_j$ cancel out. In our case, this yields the local Heisenberg Hamiltonian.

\subsection{Overview of Techniques for Electronic structure}\label{sec:overview}
We reduce from an instance of Fermi-Hubbard.
The interaction graph has an edge for every pair of fermions with a non-zero interaction term.
Given an input Hamiltonian of this form, we create a set of orbitals corresponding to the vertices in the interaction graph. $\phi_i$ is the orbital for vertex $i$. Each $\phi_i$  is a superposition of what we call {\em primitive} orbitals, which are just Gaussians centered at various points in space. For the most part, these points are spaced out from all the other points by a parameter $\Gamma$, which is set to be large. For every edge $\{i, j\}$ in the interaction graph of the Fermi-Hubbard Hamiltonian,  there is a pair of primitive orbitals, one in $\phi_i$ and one in $\phi_j$, such that the two primitive orbitals are a distance $\gamma_{i,j}$ apart. The $\gamma_{i,j}$’s are small compared to $\Gamma$.  

The dominant term that emerges from this construction
is  the kinetic energy between two Gaussians with exponent $\alpha$ that are separated by a distance of $\gamma_{i,j}$ (with a slight correction
due to the fact that  the Gaussians are not exactly pairwise orthogonal).
Each distance $\gamma_{i,j}$ can then be tuned to obtain the desired coefficient to encode the Fermi-Hubbard Hamiltonian. Each orbital also includes a primitive orbital with exponent  $\beta > \alpha$ in
order to increase the onsite-repulsion term, ensuring that the ground space for the effective Hamiltonian
has exactly one electron per spatial orbital. Thus, each of our orbitals has the form
\begin{equation}
\phi_{i} (\mathbf r) = 2^{-1/2} \phi_{i, 0}(\mathbf r) + {(2d)}^{-1/2} \sum_{i=1}^d \phi_{i, l}(\mathbf r)
,
\end{equation}
where each $\phi_{i, l}$ is a Gaussian and the parameter $d$ is an upper bound on the degree of the graph.
The functions $\phi_{i, 0}$ all have some large exponent $\beta$ and are therefore more concentrated than the functions $\phi_{i, l}$ for $l > 0$ which have a  smaller exponent $\alpha$.

We use two approximation steps
which ultimately show that the  electronic structure Hamiltonian $\ham{ES}$ closely approximates the Fermi-Hubbard Hamiltonian $\ham{Hubb}$.
\begin{equation}\label{eq:flow}
\ham{ES} 
\overset{\labelcref{sec:nonorthonormality}}{\rightarrow}
\ham{round} 
\overset{\labelcref{sec:approx}}{\rightarrow}
\ham{main}
\overset{\labelcref{sec:qma-proof}}{\propto}
\ham{Hubb},
\end{equation}
Each step introduces some small error,
the bounding of which constitutes the bulk of the technical work in our proof.

The transition  from $\ham{ES}$ to $\ham{round}$
includes two approximation steps.
The first approximation arises from
the fact that the orbitals $\boldsymbol \phi$ that we use are not perfectly orthonormal.
However, there is an orthonormal basis $\tilde{\boldsymbol \phi}$ that is very close to $\boldsymbol \phi$.
We show that the difference is sufficiently small that we can proceed with the coefficients from the nonorthonormal basis but using the elementary operators of the orthonormal basis.
There is one exception to this approximation: the overlap of the Gaussians
that are relatively close (distance $\gamma_{i,j}$ apart) has a non-negligible
effect and requires a slight correction to the corresponding kinetic
energy coefficient.
In the second approximation step, we drop the  interactions of primitive orbitals
that are at least a distance of $\Gamma$ apart, resulting in 
an expression with many fewer terms.
The effect of applying both approximations results in the Hamiltonian
$\ham{round}$.
The transition from $\ham{round}$ to $\ham{main}$ involves dropping
the potential-energy terms that involve more than one primitive orbital.
The difference between $\ham{round}$ and $\ham{main}$ is 
an energy offset which is constant for a fixed number of electrons
plus an error term which we bound in the proof. 
We then show that the parameters can be set so that the coefficients
of $\ham{main}$ approximate the Fermi-Hubbard model to within any  inverse
polynomial. 

\subsection{Product states}

Classical algorithms for finding the ground state energy of quantum Hamiltonians are often limited by the fact that the ground state seems to have no concise classical description.
For that reason, chemists often try to find the lowest-energy Slater determinant, known as the Hartree-Fock state.
Within a fixed basis $\boldsymbol \phi = (\phi_1, \ldots, \phi_n)$, a Slater determinant is a state of the form
\begin{equation}
\ket{\mathrm{SD}(B)} = b_1^{\dagger} b_2^{\dagger} \cdots b_{\eta}^{\dagger} \ket{\boldsymbol 0},
\end{equation}
where each $b_i = \sum_{j=}^{n} B_{i, j} a_j$ is a sum of annihilation operators in the original basis and the rows of the $\eta \times n$ matrix $B$ are orthonormal.

\begin{definition}[Lowest-energy Slater determinant (LESD)]
Given a local fermionic Hamiltonian in a fixed basis of size $n$, number $\eta$ of electrons, and bounds
$b > a$, where $b-a = 1 / \poly(n)$, determine whether the lowest-energy Slater determinant has energy at  most $a$ or at least $b$.
The Slater determinant is specified the matrix $\eta \times n$ $B$ with entries specified by polynomially many bits.
\end{definition}

\begin{theorem}[informal]\label{thm:LESD-NP-hard-informal}
LESD for electronic structure Hamiltonians ($\ham{ES}$ as defined in \ref{eq:es}) is NP-complete. 
\end{theorem}

Schuch and Verstraete showed that the LESD problem for generic quartic number-preserving fermionic Hamiltonians is NP-hard~\cite[arXiv version]{schuch2009computational}.
We show NP-hardness for the restricted class of such Hamiltonians with coefficients implied by a basis and external potential as in~\crefrange{eq:es}{eqn:u};
that is, our~\cref{thm:es-informal,thm:LESD-NP-hard-informal} cover the same class of electronic structure Hamiltonians and differ only in the class of states to be optimized over. 
Schuch and Verstraete's proof for the QMA-hardness of electronic structure with magnetic fields could likely be extended to the NP-hardness of the Slater determinant version, but neither we nor they have done so.

\subsection{A Note on Notation}

By $[n]$, we mean the set $\{1, 2, \ldots, n\}$.
We use $\|\cdot\|$ for the spectral norm of a matrix and the Euclidean norm of a vector.
We use $|\cdot|$ for the element-wise scalar norm.

\section{Fermi-Hubbard Model}
\label{sec:fermihubb}

We will show that the version of the Fermi-Hubbard Hamiltonian
problem described below is QMA-complete. 
In order for the Fermi-Hubbard model to approximate the
antiferromagnetic Heisenberg from which we are reducing,
we need a large onsite-repulsion term $u_0$
to penalize orbitals with double occupancy. 
The Fermi-Hubbard problem remains QMA-complete for any
$u_0$ that satisfies the lower bound in the theorem stated below.
For the reduction from Fermi-Hubbard to Electronic Structure,
we require that the $\ham[t]{Hubb}_{i,j}$ coefficients
are bounded by a polynomial in $n$, the number of electrons.
The hardness  result that we prove establishes that Fermi-Hubbard
remains hard, even under that constraint. 

\begin{theorem}[QMA-completeness of Hubbard Hamiltonian with  uniform onsite repulsion]
\label{thm:hubbard}
There exist constants $p > q > 0$ such that for all $\ham[u]{Hubb}_0 \geq n^{14 + 3 p + 2q}$, determining to precision $n^{-q}$ the ground state energy in the $n$-particle subspace of a Hubbard Hamiltonian
\begin{equation}
\ham{Hubb}
=
\ham[u]{Hubb}_0 \sum_{i \in [n]}
n_{i, +1} n_{i, -1}
+
\sum_{i < j}
\ham[t]{Hubb}_{\substack{i, j \\ \sigma \in \{\pm 1\}}}
\left(
a_{i, \sigma}^{\dagger}
a_{j, \sigma}
+
a_{j, \sigma}^{\dagger}
a_{i, \sigma}
\right)
\end{equation}
subject to $\left|\ham[t]{Hubb}_{i, j}\right| \leq \sqrt{n^p \ham[u]{Hubb}_0}$
is QMA-complete.
\end{theorem}

We will reduce from the antiferromagnetic Heisenberg Hamiltonian problem:

\begin{definition}[Antiferromagnetic Heisenberg Hamiltonian]
An instance of antiferromagnetic Heisenberg Hamiltonian is defined by an edge-weighted graph $G = (V, E)$ with $\kappa: E \mapsto \mathbb R_{\geq 0}$ as
\begin{equation}
H^{\mathrm{(Heis)}}(G, w)
=
\sum_{
\{i, j\} \in E} \kappa_{i,j} \left(X_i X_j + Y_i Y_j + Z_i Z_j\right)
\end{equation}
\end{definition}

We will require in our reduction that the coefficients $\kappa_{i,j}$
are bounded by a polynomial in the number of qubits.
Although not explicitly stated,
the following theorem is proven in ~\cite{piddock2017complexity}.

\begin{theorem}[QMA-completeness of antiferromagnetic Heisenberg Hamiltonian~\cite{piddock2017complexity}]
\label{thm:heisenberg}
Finding the ground state of an antiferromagnetic Heisenberg Hamiltonian is QMA-complete even when restricted to families of Hamiltonians in which
the coefficients are bounded by a polynomial in the number of qubits.
\end{theorem}

To prove~\cref{thm:hubbard}, we show that for sufficiently large $\ham[u]{Hubb}_0$, the Hubbard model approximates an antiferromagnetic Heisenberg model up to second order in perturbation theory.

We'll treat $\ham[U]{Hubb} = \ham[u]{Hubb}_0 \sum_i n_{i, +1} n_{i, -1}$ as the penalty term and $\ham[T]{Hubb} = \ham{Hubb} - \ham[U]{Hubb}$ as the perturbation.
To convert the fermionic Hamiltonians above to qubit Hamiltonians, we use the Jordan-Wigner transform with the ordering $(1, +1)$, $(1, -1)$, $(2, +1)$, $(2, -1)$, \ldots.
For the full Hilbert space $\mathcal H$ we'll use a basis of one qubit per spin orbital.
For the ground space $\mathcal H_0$ of $\ham[U]{Hubb}$, we'll use a basis of one qubit per spatial orbital, the latter spanning the half-filled subspace of the corresponding pair of spin orbitals.
We associate the occupancy of the orbitals of spin $+1$ and $-1$ with the qubit states $\ket{0}$ and $\ket{1}$, respectively.
Let $\Pi_0$ be the projector onto $\mathcal H_0$ and $\Pi_1 = I - \Pi_0$ the projector onto the orthogonal subspace.
In $\mathcal H_0$, $\ham[U]{Hubb}$ is zero ($\ham[U]{Hubb}_0 = \Pi_0 \ham[U]{Hubb} \Pi_0 = 0$), and outside it is at least $\ham[u]{Hubb}_0$.
In the half-filling regime, the ground space of $\ham[U]{Hubb}$ is spanned by those basis states having exactly one electron in each spatial orbital.
In $\mathcal H_0$, $\ham[T]{Hubb}$ vanishes.
In the notation below, we use a bit to indicate whether an orbital is filled.
For edge $\{i,j\}$, the first two bits correspond to orbitals
$\phi_{i,+1}$ and $\phi_{i,-1}$ and the last two bits correspond to
$\phi_{j,+1}$ and $\phi_{j,-1}$. So the state $\ket{0110}$ has
$\phi_{i,-1}$ and $\phi_{j,+1}$ filled.
The ``excitation'' terms are
\begin{align}
\ham[T]{Hubb}_{1, 0}
&=
\Pi_1 \ham[T]{Hubb} \Pi_0
=
\sum_{\{i, j\} \in E}
    {(-1)}^{j - i - 1}
    \ham[t]{Hubb}_{i, j}
    {\left[
        \left(
        \ket{1100} + \ket{0011}
        \right)
        \left(
        \bra{1001} - \bra{0110}
        \right)
        \right]}_{i, j}
.
\end{align}
With this, using~\cref{thm:pert} from the next section, we get
\begin{align}
H^{(\mathrm{eff})}
&=
- 
\ham[T]{Hubb}_{0, 1}
{\left[
\ham[U]{Hubb}_1
\right]}^{-1}
\ham[T]{Hubb}_{1, 0}
=
\sum_{\{i, j\} \in E}
\frac{
2
{\left(
\ham[t]{Hubb}_{i, j} 
\right)}^2
}{
\ham[u]{Hubb}_0
}
\left(W_{i, j} - 1\right)
=
c_{\mathrm{eff}}
+
\sum_{\{i, j\} \in E}
\ham[h]{eff}_{i, j}
W_{i, j}
,
\end{align}
where
\begin{align}
\ham[h]{eff}_{i, j}
&=
2
\frac{
{\left(
\ham[t]{Hubb}_{i, j}
    \right)
}^2
}{
\ham[u]{Hubb}_0
}
,
&
c_{\mathrm{eff}}
&=
-
\frac{1}{\ham[u]{Hubb}_0}
\sum_{\{i, j\} \in E}
{\left(
\ham[t]{Hubb}_{i, j}
    \right)
}^2
.
\end{align}

\subsection{Perturbation Theory}

We will use the following formulation of second-order perturbation theory, adapted from a special case of the more general formulation by Bravyi et al.~\cite{bravyi2011schrieffer}.
\begin{theorem}\label{thm:pert}[Second-order perturbation theory]
Consider a Hamiltonian $H = \ham{pen} + \ham{pert}$.
Let $\Pi_0$ be the projector onto the ground space of $\ham{pen}$, and $\Pi_1 = 1 - \Pi_0$.
Define
\begin{equation}
\ham{eff}
=
+
\ham{pert}_0 - \ham{pert}_{0, 1} {\left(\ham{pen}\right)}^{-1} \ham{pert}_{1, 0}
,
\end{equation}
where $A_{i} = \Pi_i A \Pi_i$ and $A_{i, j} = \Pi_i A \Pi_j$.
If $\ham{pen}_0 = 0$ and $\ham{pen}_1 \geq \Delta \geq 2 \ham{pert}$, then
\begin{equation}
\norm{
H_{\mathrm{low}}
-
\ham{eff}
}
\leq
O \left(
\frac{
\norm{\ham{pert}}^3
}{
\Delta^2
}
\right)
,
\end{equation}
where $H_{\mathrm{low}}$ is the projection of $H$ onto its eigenspace with eigenvalues at most $\Delta / 2$.
\end{theorem}

\subsection{Fermi-Hubbard is QMA-Hard}

\begin{proof}[Proof of~\cref{thm:hubbard}]

There are constants $p, q \geq 0$ such that it is QMA-hard to find the ground state energy to precision $n^{-q}$ of 
\begin{align}
\ham{Heis}
&=
\sum_{\{i, j\} \in E}
\kappa_{i, j} W_{i, j}
\end{align}
subject to $0 \leq \kappa_{i, j} \leq n^{p}$.
Consider such an instance.
We want to choose $\ham[u]{Hubb}_0$ and $\ham[t]{Hubb}_{i, j}$ such that
\begin{equation}
\ham{Heis}
=
\ham{eff} - c_{\mathrm{eff}}
\label{eq:Heis-eff-equal}
\end{equation}
and
\begin{equation}
\norm{\ham{Hubb}_{\mathrm{low}} - \ham{eff}}
=
o(n^{-q})
\label{eq:Heis-eff-error}
.
\end{equation}
The first constraint, \cref{eq:Heis-eff-equal}, is
\begin{equation}
\kappa_{i, j}
=
\ham[h]{eff}_{i, j}
=
2
\frac{
{\left(
\ham[t]{Hubb}_{i, j}
    \right)
}^2
}{
\ham[u]{Hubb}_0
}
\end{equation}
or
\begin{equation}
\ham[t]{Hubb}_{i, j}
=
\pm
\sqrt{
\ham[u]{Hubb}_0 \kappa_{i, j} / 2
}
.
\end{equation}
Therefore, for any $\kappa_{i, j}$ such that $\left|\kappa_{i, j}\right| \leq n^p$
    we can choose $\ham[t]{Hubb}_{i, j}$ such that $\left|\ham[t]{Hubb}_{i, j}\right| \leq \sqrt{n^p \ham[u]{Hubb}_0}$ and that~\cref{eq:Heis-eff-equal} is satisfied.
To satisfy the second constraint,~\cref{eq:Heis-eff-error}, we use second-order perturbation theory (\cref{thm:pert}).

Furthermore, the assumption that $\ham[u]{Hubb}_0 \geq n^{14 + 3p + 2q}$ implies that the condition of ~\cref{thm:pert} is met:
\begin{align}
\norm{\ham[T]{Hubb}}
&
\leq
\sum_{\substack{\{i, j\} \in E \\ \sigma \in \{\pm 1\}}}
\left|
\ham[t]{Hubb}_{i, j}
\right|
\\
&
\leq
\underbrace{n^2}_{\{i, j\}, \sigma} \cdot 
\underbrace{\sqrt{\ham[u]{Hubb}_0 n^p}}_{\ham[t]{Hubb}_{i, j}}
\label{eq:T-Hubb-UB}
\\
&=
\sqrt{\ham[u]{Hubb}_0}
\sqrt{n^{4 + p}}
\\
&\leq
\sqrt{\ham[u]{Hubb}_0}
\cdot
\frac12
\sqrt{n^{14 + 3p + 2q}}
& n \geq 2; p, q \geq 0
\\
&\leq
\sqrt{\ham[u]{Hubb}_0}
\cdot
\frac12
\sqrt{\ham[u]{Hubb}_0}
=
\frac12 \ham[u]{Hubb}_0
.
\end{align}
\cref{thm:pert} then yields
\begin{align}
\norm{
\ham{Hubb}_{\mathrm{low}}
-
\ham{eff}
}
&\leq 
O\left(
\frac{
\norm{\ham[T]{Hubb}}^3
}{
{\left(
\ham[u]{Hubb}_0
\right)}^2
}
\right)
\leq
O\left(
\frac{
\overbrace{n^6 n^{1.5 p} {\left(\ham[u]{Hubb}_0\right)}^{1.5}}^{\text{~\labelcref{eq:T-Hubb-UB}}}
}{
{\left(\ham[u]{Hubb}_0\right)}^{2}
}
\right)
\\
&=
O\left(
\frac{
n^6 n^{1.5 p}
}{
\sqrt{\ham[u]{Hubb}_0}
}
\right)
\leq
O\left(
\frac{
n^6 n^{1.5 p}
}{
\sqrt{n^{14+3p+2q}}
}
\right)
=
O\left(
n^{-(q+1)}
\right)
=
o\left(
n^{-q}
\right)
.
\end{align}
\end{proof}

\section{Electronic structure}
\label{sec:ES}
Here we prove our main result:
\begin{customthm}{\labelcref*{thm:es-informal}}[QMA-completeness of electronic structure in fixed basis set]\label{thm:electronic-structure}
Determining the ground state energy of an electronic structure Hamiltonian in a fixed basis set and with fixed particle number to inverse-polynomial precision is QMA-complete.
\end{customthm}
We'll start by defining a set of $n$ spatial orbitals. Once the orbitals are fixed,
the $t$ and $u$ coefficients are determined by the integrals in (\ref{eqn:t}) and (\ref{eqn:u}),
which then yields  the physical Hamiltonian
\begin{equation}
\label{eq:phys}
\ham{ES} = T + U = 
\sum_{\substack{i, j \in [n] \\ \sigma \in \{\pm 1\}}} t_{i, j} a_{i, \sigma}^{\dagger} a_{j, \sigma}
+
\frac12
    \sum_{\mathclap{
        \substack{i, j, k, l \in [n] \\ \sigma, \tau \in {\{\pm 1\}}}
    }}
u_{i, j, k, l}
a_{i, \sigma}^{\dagger}
a_{j, \tau}^{\dagger}
a_{k, \tau}
a_{l, \sigma}
\end{equation}
in the absence of any external potential ($V = 0$).
We show that,
when restricted to the subspace with exactly $n$ electrons (with arbitrary spin),
this yields an effective Hamiltonian that is close, up to rescaling and shifting, to a Fermi-Hubbard Hamiltonian
\begin{equation}
\ham{Hubb}
=
\ham[u]{Hubb}_0 \sum_{i \in [n]}
n_{i, +1} n_{i, -1}
+
\sum_{i < j}
\ham[t]{Hubb}_{i, j}
\left(
a_{i, \sigma}^{\dagger}
a_{j, \sigma}
+
a_{j, \sigma}^{\dagger}
a_{i, \sigma}
\right)
\label{eq:fermihubb}
\end{equation}
with where there are constants $p > q > 0$ such that $\ham[u]{Hubb}_0 \ge n^{14+3p+2q}$ and $\left|\ham[t]{Hubb}_{i, j}\right| \leq \sqrt{n^p \ham[u]{Hubb}_0}$ for all edges $\{i,j\}$.

\subsection{Orbitals}\label{sec:orbitals}
Recall from~\cref{sec:overview} that our goal is a basis of orbitals such that the electronic structure Hamiltonian in that basis is sufficiently close to a Hubbard Hamiltonian. We define here a set of orbitals that effectively encodes the interaction graph of the Hubbard Hamiltonian. Each orbital represents a vertex of the interaction graph and
consists of  a superposition of Gaussians centered at various points in space. For the most part, these Gaussians are far apart from each other. 
If two vertices are connected by an edge, then their corresponding orbitals
have two Gaussians that are relatively close to each other. This distance between the Gaussians can
be tuned to match the interaction coefficient in the Hubbard Hamiltonian.
Let
\begin{equation}
\xi_{\alpha}(\mathbf r) = 
{\left(\frac{2\alpha}{\pi}\right)}^{3/4}
\exp\left(-\alpha \norm{\mathbf r}^2\right)
\end{equation}
be the Gaussian centered at $\mathbf 0 \in \mathbb R^3$ with exponent $\alpha > 0$.
Each of our orbitals will be a superposition of Gaussians.
The centers of these Gaussians will be a set of points ${\left\{\mathbf x_{i,l}\right\}}_{i, l}$
in $\mathbb R^3$, where $i \in [n]$ and $l \in \{0\} \cup [d]$,
and
$d \leq n-1$ is an upper bound on the maximum degree of the interaction graph $G$.
Note that although the points are in $\mathbb R^3$, the properties we require of them can be satisfied by placing them all along a line; 
we'll use an arrangement in the 2-dimensional plane for convenience.
We require two properties of this set of points:
\begin{enumerate}
    \item For each edge $\{i, j\}$ in the interaction graph, there is exactly one pair $(l, l') \in {[d]}^2$ such that 
        $\norm{\mathbf x_{i, l} - \mathbf x_{j, l'}} = \gamma_{i, j} > 0$.
        Let $\gamma_{\min}$ and $\gamma_{\max}$ be lower and upper bounds on $\gamma_{i, j}$ over $\{i, j\} \in E$.
    \item Every other pair of points is at least $\Gamma \gg \gamma_{\max}$ apart (in Euclidean distance).
\end{enumerate}
The important part is the $m = |E|$ pairs of points such that points from different pairs are at least a distance of $\Gamma$ apart.
Each pair of points is associated with an edge $\{i,j\}$ in the
interaction graph.
The pair of points associated with edge $\{i,j\}$ will be $\gamma_{i,j}$ apart,
where $\Gamma \gg \gamma_{i,j}$.
In addition, there is a set 
$\orbitalcenters$ of $(d+1)n - 2m$ points each of which is a distance at least
$\Gamma$ from any other point in the construction.
The points associated with vertex $i$ in the
interaction graph are as follows:
\begin{enumerate}
    \item $x_{i,0}$ is a point from $\orbitalcenters$.
    \item If $p \le \degree(i)$ and $j$ is the $p$-th neighbor of vertex $i$, then $x_{i,p}$ will be one of the points from the pair associated with edge $\{i,j\}$. (The other point from the pair will belong to vertex $j$.)
    \item If $p > \degree(i)$, then $x_{i,p}$ is a point from $\orbitalcenters$.
(These are just dummy neighbors to ensure that all of the orbitals have the same form.)
\end{enumerate}
With these points, we can define the {\em primitive} orbitals
\begin{equation}
\phi_{i, p}(\mathbf r) = 
\begin{cases}
\xi_{\beta}\left(\mathbf r - \mathbf x_{i, 0}\right) & p =0, \\
\xi_{\alpha}\left(\mathbf r - \mathbf x_{i, p}\right), & \text{otherwise},
\end{cases}
\end{equation}
where $\alpha$ and $\beta$ are positive constants to be set later.
Ultimately, we will need $\beta \gg \alpha$.
The {\em composite} orbitals that we'll use in the construction will be superpositions of these primitive orbitals:
\begin{equation}\label{eq:orbital-def}
\phi_i(\mathbf r) = 
\frac1{\sqrt{2}} \phi_{i, 0}(\mathbf r) + 
\frac1{\sqrt{2 d}} \sum_{l=1}^{d} \phi_{i, l}(\mathbf r)
.
\end{equation}
It will be convenient to be able to refer to the indices of the primitive orbitals that are a distance
$\gamma_{i,j}$ apart, corresponding to edge $\{i,j\}$. Define $\block(i,j) = \{(i,p),(j,q)\}$,
where $j$ is the $p$-th neighbor of $i$ and $i$ is the $q$-th neighbor of $j$.

We will eventually show that
the kinetic energy terms between the primitive orbitals that are separated
by only a distance of $\gamma_{i,j}$ will be the dominant terms in the Hamiltonian (besides the onsite-repulsion).
We will then tune the $\gamma_{i,j}$ distances so that the coefficients
resulting from kinetic energy integrals scale with the $\ham[t]{Hubb}_{i,j}$ from \cref{eq:fermihubb},
which are the coefficients in the Fermi-Hubbard Hamiltonian from which
we are reducing. The radius $\beta$ will be chosen to be large
enough so that the potential energy coefficients $u_{i,i,i,i}$
effectively result in a $u_0 n_{i,+1} n_{i,-1}$ with a large
coefficient $u_0$. 

\begin{figure}[htbp]
\centering
\includegraphics[width=2.5in]{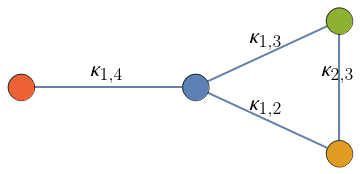}
\\
\vspace{.3in}

\includegraphics[width=\textwidth]{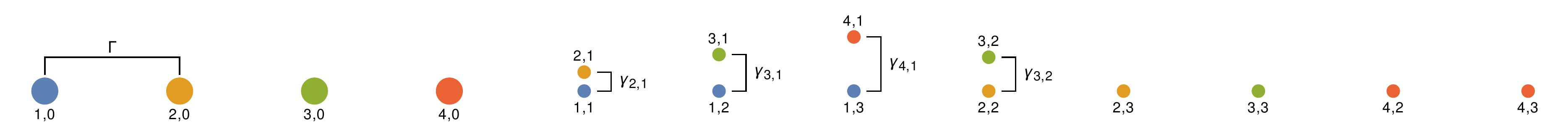}
  
\caption{
The top figure is the interaction graph of a Heisenberg Hamiltonian.
The bottom figure shows a possible placement of the primitive orbitals.
Larger points represent Gaussians with radius $\beta$.
Smaller points represent Gaussians with radius $\alpha$.
The orbitals are color coded according to which vertex they belong to from the interaction graph, and thus which composite orbital they contribute to..
For example, the orbital associated with the blue vertex in the interaction graph would be a superposition of the blue Gaussians in the bottom figure.
The amplitude of the large blue Gaussian on the left is $1/\sqrt{2}$.
The amplitude of each of the smaller blue Gaussians is $1/\sqrt{2d}$, where here $d = 3$.}\label{fig:orbitals}
\end{figure}

The construction is illustrated in Figure \ref{fig:orbitals} with a small example.
The orbitals are strictly positive everywhere, so the overlap $\int d \mathbf r \phi_i^*(\mathbf r) \phi_j(\mathbf r)$ cannot be exactly zero, but we will show that it's very close.
That is, we will show that the orbitals are not perfectly orthonormal, but that they are sufficiently close.
For now, we'll proceed as if they are, and address the effect of the nonorthonormality in \cref{sec:nonorthonormality}.
The purpose of including the $\phi_{i, 0}$ component
as part of the orbital, which is far away from every other primitive orbital center, is to decouple the scale of the onsite-repulsion term in the Hamiltonian from that of interaction term, which are effected by the one of the components $\phi_{i, l}$ for each orbital $\phi_i$.
To this end, we will ultimately set $\beta \gg \alpha$.
Including the other components ${\left\{\phi_{i, l'}\right\}}_{l' > \degree(i)}$ is simply to ease the analysis by making all of the orbitals ${\left\{\phi_i\right\}}_i$ have integrals, over single-electron operators, that are of approximately the same form.

\subsubsection{Integrals of Operators over Gaussians}

Since the composite orbitals are superpositions of primitive orbitals, the
expressions for overlap, kinetic energy, and potential energy for the composite
orbitals will be linear combinations of the corresponding expression for combinations  of  primitive orbitals.
The following integrals of operators over Gaussians will be useful in expressing these
terms for the primitive orbitals.

The overlap of two Gaussians with exponents $\alpha$ and $\beta$ with centers $\mathbf x$ apart:
\begin{equation}
s_{\alpha, \beta}(\norm{\mathbf x})
=
\int d \mathbf r 
\xi_{\alpha} (\mathbf r)
\xi_{\beta} (\mathbf r - \mathbf x)
=
{\left( 
    \frac{2 \sqrt{\alpha \beta}}{\alpha + \beta}\
\right)}^{3/2}
\exp
\left(-\frac{\alpha \beta}{\alpha + \beta} \norm{\mathbf x}^2\right).
\end{equation}
Due to the rotational invariance of the Gaussians, all of the functions defined in this subsection
depend only on the magnitude of their argument, and so we will write, for example, $s\left(\norm{\mathbf x}\right)$.

We can define the $n(d+1) \times n(d+1)$ matrix $S$ of overlap between the primitive orbitals, where each row and column is indexed by a pair $(i,p)$ corresponding to a primitive orbital:
\begin{equation}
s_{(i,p),(j,q)} = -\frac12 \int d \mathbf r \phi^*_{(i,p)}(\mathbf r)  \phi_{(j,q)}(\mathbf r)
\label{eqn:s-matrix-def}
\end{equation}
Note that $s_{\alpha,\beta}(\norm{\mathbf x})$ denotes the overlap of primitive  orbital $\phi_{i,0}$ (whose exponent is $\beta)$ and $\phi_{j,p>0}$ (whose exponent is $\alpha$), where $\phi_{i,0}$ and $\phi_{j,p}$
are separated by a distance $\norm{\mathbf x}$: $s_{(i,0),(j,p)} = s_{\alpha,\beta}(\norm{\mathbf x})$.
The overlap of two primitive orbitals with the same exponent are denoted by:
\begin{align}
s_{\alpha} (\norm{\mathbf x})
& =
s_{\alpha, \alpha} (\norm{\mathbf x})
=
\exp\left(- \alpha \norm{\mathbf x}^2 / 2\right),\\
s_{\beta} (\norm{\mathbf x})
& =
s_{\beta, \beta} (\norm{\mathbf x})
=
\exp\left(- \beta \norm{\mathbf x}^2 / 2\right).
\end{align}
Therefore $s_{(i,0),(j,0)} = s_{\beta} (\norm{\mathbf x})$, where primitive orbitals $\phi_{(i,0)}$ and $\phi_{(j,0)}$
are a distance $\norm{\mathbf x}$ apart. Also, $s_{(i,p),(j,q)} = s_{\alpha} (\norm{\mathbf x})$, where $p,q >0$ and primitive orbitals $\phi_{(i,p)}$ and $\phi_{(j,q)}$
are a distance $\norm{\mathbf x}$ apart.

The kinetic energy between two Gaussians with exponents $\alpha$ and $\beta$ with centers $\mathbf x$ apart is:
\begin{equation}
t_{\alpha, \beta}(\norm{\mathbf x}) 
=  
-\frac12
\int d \mathbf r 
\xi_{\alpha} (\mathbf r)
\nabla^2
\xi_{\beta} (\mathbf r - \mathbf x)
=
2^{3/2} 
\frac{{(\alpha \beta)}^{7/4}}{{(\alpha + \beta)}^{5/2}}
\left(3 - 2 \mu \norm{\mathbf x}^2\right)
\exp\left(-\mu \norm{\mathbf x}^2\right)
\leq
\frac32 \max \left\{\alpha, \beta\right\}
,
\label{eq:gen-t-integral}
\end{equation}
where $\mu = \alpha \beta /(\alpha + \beta)$,
with
\begin{align}
t_{\alpha}(\norm{\mathbf x}) = t_{\alpha, \alpha}(\norm{\mathbf x})
& =
\frac{\alpha}2
\left(3 - \alpha \norm{\mathbf x}^2\right)
\exp\left(-\alpha \norm{\mathbf x}^2 / 2\right),
\label{eq:t-alpha-def}\\
t_{\beta}(\norm{\mathbf x}) = t_{\beta, \beta}(\norm{\mathbf x})
& =
\frac{\beta}2
\left(3 - \beta \norm{\mathbf x}^2\right)
\exp\left(-\beta \norm{\mathbf x}^2 / 2\right).
\label{eq:t-beta-def}
\end{align}
 Define $T$
to be the $n(d+1) \times n(d+1)$ matrix of kinetic energy terms between primitive orbitals. An entry
of matrix $T$ is
\begin{equation}
t_{(i,p),(j,q)} = -\frac12 \int d \mathbf r \phi^*_{(i,p)}(\mathbf r) \nabla^2 \phi_{(j,q)}(\mathbf r).
\label{eqn:t-matrix-def}
\end{equation}
Therefore, $t_{(i,0),(j,0)} = t_{\beta} (\norm{\mathbf x})$, where primitive orbitals $\phi_{(i,0)}$ and $\phi_{(j,0)}$
are a distance $\norm{\mathbf x}$ apart. Also, $t_{(i,p),(j,q)} = t_{\alpha} (\norm{\mathbf x})$, where $p,q >0$ and primitive orbitals $\phi_{(i,p)}$ and $\phi_{(j,q)}$
are a distance $\norm{\mathbf x}$ apart.

The potential integrals:
\begin{align}
\ham[u]{Coul}_{\alpha}(\norm{\mathbf x})
&=
\int d\mathbf r d \mathbf s
{\xi_{\alpha}(\mathbf r)}^2
{\xi_{\alpha}(\mathbf s - \mathbf x)}^2
\norm{\mathbf r - \mathbf s}^{-1}
=\sqrt{\frac{4 \alpha}{\pi}}
F_0\left(\alpha \norm{\mathbf x}^2\right)
\leq 2 \sqrt{\alpha}
,
\label{eq:coulomb-integral}
\\
\ham[u]{exch}_{\alpha}(\norm{\mathbf x})
&=
\int d\mathbf r d \mathbf s
\xi_{\alpha}(\mathbf r)
\xi_{\alpha}(\mathbf r - \mathbf x)
\xi_{\alpha}(\mathbf s)
\xi_{\alpha}(\mathbf s - \mathbf x)
\norm{\mathbf r - \mathbf s}^{-1}
=
\exp\left(- \alpha \norm{\mathbf x}^2\right) \ham[u]{Coul}_{\alpha}(0)
,
\label{eq:exch-integral}
\\
\ham[u]{other}_{\alpha}(\norm{\mathbf x})
&=
\int d\mathbf r d \mathbf s
{\xi_{\alpha}(\mathbf r)}^2
\xi_{\alpha}(\mathbf s)
\xi_{\alpha}(\mathbf s - \mathbf x)
\norm{\mathbf r - \mathbf s}^{-1}
=
\exp\left(- \alpha \norm{\mathbf x}^2 / 2\right) \ham[u]{Coul}_{\alpha}\left(\mathbf x / 2\right)
,
\end{align}
where
\begin{equation}
F_k(x) = \int_0^1 e^{-x t^2} t^{2k} dt
\end{equation}
is the Boys function of order $k$.
The analogous definitions for $\ham[u]{other}_{\beta}(\mathbf x)$, $\ham[u]{other}_{\beta}(\mathbf x)$,
and $\ham[u]{other}_{\beta}(\mathbf x)$ use Gaussians with exponent $\beta$.
The potential energy terms on the primitive orbitals are represented by an $n^2(d+1)^2 \times n^2(d+1)^2$ matrix
$U$, where each row and column is indexed by a pair of primitive orbitals $[(i,p),(j,q)]$. The entry in
row $[(i,p),(j,q)]$ and column $[(k,r),(l,s)]$ is the potential energy term for orbitals
$\phi_{(i,p)}(\mathbf r)$, $\phi_{(j,q)}(\mathbf r)$, $\phi_{(k,r)}(\mathbf r)$, and $\phi_{(l,s)}(\mathbf r)$:
\begin{equation}
u_{[(i,p),(j,q)],[(k,r),(l,s)]}
=
\int d \mathbf r d \mathbf s
\frac{
\phi^*_{(i,p)}(\mathbf r) \phi^*_{(j,q)}(\mathbf s)
\phi_{(k,r)}(\mathbf s) \phi_{(l,s)}(\mathbf r)
}{\left|\mathbf r - \mathbf s\right|}.
\label{eqn:u-matrix-def}
\end{equation}
The definitions of the integral functions above correspond to the situation where all four indices
$(i,p)$, $(j,q)$, $(k,r)$, and $(l,s)$ denote at most two distinct orbitals with the same exponent.
Specifically, for $p,q > 0$, where $\norm{\mathbf x}$ is the distance between $\phi_{(i,p)}$ and
$\phi_{(j,q)}$:
\begin{align}
    u_{[(i,p),(j,q)],[(j,q),(i,p)]} &= \ham[u]{Coul}_{\alpha}(\mathbf x),\\
    u_{[(i,p),(j,q)],[(i,p),(j,q)]} &= \ham[u]{exch}_{\alpha}(\mathbf x),\\
    u_{[(i,p),(i,p)],[(i,p),(j,q)]} &= \ham[u]{other}_{\alpha}(\mathbf x).
\end{align}

\subsection{Orthonormalizing and rounding}\label{sec:nonorthonormality}
Having constructed our orbitals, we now make two approximations  to get a clean, ``round'' Hamiltonian $\ham{round}$.
First, the orbitals we defined in~\cref{eq:orbital-def} are slightly nonorthonormal, and so we derive a related orthonormalized basis in which the electronic structure Hamiltonian doesn't change too much.
Second, we remove contributions to the Hamiltonian from the electron-electron interaction pairs of primitive orbitals that are far ($>\Gamma$) away from each other.
The error of these approximations is quantified by~\cref{lem:RO-error}.

The matrix S is defined in (\ref{eqn:s-matrix-def}) to be the overlap matrix of the \emph{primitive} orbitals.
We can construct a set of orthonormal primitive orbitals by setting:
\begin{equation}
\tilde{\phi}_{i, k} (\mathbf r)= \sum_{j, l} {\left[S^{-1/2}\right]}_{(j, l), (i, k)} \phi_{j, l}(\mathbf r)
\end{equation}
and new orthonormal composite orbitals
\begin{equation}
\tilde{\phi}_i(\mathbf r) = 
\frac1{\sqrt{2}} \tilde{\phi}_{i, 0}(\mathbf r) + 
\frac1{\sqrt{2 d}} \sum_{l=1}^{d} \tilde{\phi}_{i, l}(\mathbf r)
\end{equation}
with annihilation operators $\tilde{a}_{i, \sigma}$~\cite{helgaker2000molecular}.
The Hamiltonian in this orthonormal basis is
\begin{align}\label{eq:orthonormal-ham}
\ham{ES}
&=
\ham[T]{ES}
+
\ham[U]{ES}
=
\sum_{\substack{i, j \in [n] \\ \sigma \in \{\pm 1\}}} 
\tilde{t}_{i, j} 
\tilde{a}_{i, \sigma}^{\dagger}
\tilde{a}_{j, \sigma}
+
\frac12
\sum_{\mathclap{
\substack{i, j, k, l \in [n] \\ \sigma, \tau \in {\{\pm 1\}}}
}}
\tilde{u}_{i, j, k, l}
\tilde{a}_{i, \sigma}^{\dagger}
\tilde{a}_{j, \tau}^{\dagger}
\tilde{a}_{k, \tau}
\tilde{a}_{l, \sigma}
\end{align}
where
\begin{align}
\tilde{t}_{i, j}
&=
\int d\mathbf r \tilde{\phi}_i^* (\mathbf r) T \tilde{\phi}_j (\mathbf r)
,
\\
\tilde{u}_{i, j, k, l}
&=
\int d\mathbf r d\mathbf s
\tilde{\phi}_i^*(\mathbf r)
\tilde{\phi}_j^*(\mathbf s)
U
\tilde{\phi}_k(\mathbf s)
\tilde{\phi}_l(\mathbf r)
.
\end{align}
The matrices $\tilde{T}$ and $\tilde{U}$ showing the kinetic and potential energies using the orthonormalized primitive orbitals are analogous to the definitions (\ref{eqn:t-matrix-def}) and (\ref{eqn:u-matrix-def}). The pair $T$ and $\tilde{T}$ and the pair $U$ and $\tilde{U}$
are related by conjugation by $S^{-1/2}$:
\begin{align}
    \tilde{T} &= S^{-1/2}~ T~ S^{-1/2},\\
    \tilde{U} &= (S^{-1/2} \otimes S^{-1/2})~ U~ (S^{-1/2} \otimes S^{-1/2}).
    \label{eqn:tilde-relation}
\end{align}
Since $S \approx I$, the coefficients $\tilde{t}$ and $\tilde{u}$ for the orthonormalized orbitals
are close to $t$ and $u$ for the non-orthonormalized orbitals, but the difference needs to be carefully bounded.
We will approximate $\ham[H]{ES}$ by the Hamiltonian $\ham{round}$ that  uses
creation and annihilation operators of the orthonormal basis $\tilde{\boldsymbol \phi}$ with the original coefficients, subject to two modifications. 
First, we add a first-order correction to the off-diagonal kinetic coefficients.
Second, we remove contributions from pairs of primitive orbitals that are at least $\Gamma$ apart (which makes many terms vanish completely).

The rounded Hamiltonian is
\begin{align}
\ham{round} 
&=  
\ham[T]{round} + \ham[U]{round}.
\end{align}
The rounded kinetic operator is
\begin{align}
\ham[T]{round}
&=
\ham[t]{round}_{i, i}
\sum_{\substack{i \in [n] \\ \sigma \in \{\pm 1\}}} 
\tilde{n}_{i, \sigma}
+
\sum_{\substack{\{i, j\} \in E \\ \sigma \in \{\pm 1\}}} 
\ham[t]{round}_{i, j}
\left(
\tilde{a}_{i, \sigma}^{\dagger}
\tilde{a}_{j, \sigma}
+
\tilde{a}_{j, \sigma}^{\dagger}
\tilde{a}_{i, \sigma}
\right)
,
\\
\ham[t]{round}_{i, i} &= c_T = 
\frac12 \left(
t_{\alpha}(0) + t_{\beta}(0)
\right)
,
\label{eq:t-round-def-on}
\\
\ham[t]{round}_{i, j} 
&=
-\frac{\alpha}{4d} \sqrt{f(\omega_{i, j})}
,
\label{eq:t-round-def-off}
\end{align}
where 
\begin{align}
\omega_{i, j} &= \alpha \gamma_{i, j}^2,
&
f(\omega) &= \omega^2 \exp(-\omega).
\end{align}
Before getting to the rounded potential operator, let's consider the difference between $\ham[T]{round}$ and the true kinetic operator $\ham[T]{ES}$ in~\cref{eq:orthonormal-ham}.
Let $\psi_0 = 1 / \sqrt{2}$ and $\psi_l = 1 / \sqrt{2d}$ for $l > 0$.
Since the composite orbitals are superpositions of the primitive orbitals, the
kinetic energy term for a pair of composite orbitals is just a linear combination of 
kinetic energy terms for pairs of primitive orbitals:
\begin{align}\label{eq:t-breakdown}
&
\tilde{t}_{i, j}
=
\sum_{p, q}
\psi_{p} \psi_{q}
\tilde{t}_{(i, p), (j, q)}
.
\end{align}
We will eventually show that the kinetic energy contribution for pairs of primitive orbitals that are
at least $\Gamma$ apart will be negligible. Therefore, the only kinetic energy terms that
contribute significantly to the sum above are $\tilde{t}_{(i,p),(i,p)}$ and $\tilde{t}_{(i,p),(j,q)}$,
where $\{i,j\}$ is an edge and $\block(i,j) = \{(i,p),(j,q)\}$. 
This means that for edge $\{i,j\}$, there is only one significant term in the 
sum for $\tilde{t}_{i, j}$. If $\{i,j\}$ is not  an edge, then all of the primitive orbitals
for composite orbitals $i$ and $j$ are at least $\Gamma$ apart, and $\tilde{t}_{i, j} \approx 0$.
For the diagonal terms $\tilde{t}_{i, i}$, there are $d+1$ significant terms in the sum, corresponding
to $\tilde{t}_{(i,p),(i,p)}$ terms.
Thus, we will show that
\begin{align}
\tilde{t}_{i, j}
=
\sum_{p, q}
\psi_{p} \psi_{q}
\tilde{t}_{(i, p), (j, q)}
\approx
\begin{cases}
\sum_{p}
\psi_{p}^2
\tilde{t}_{(i, p), (i, p)}
,
& i= j,
\\
\frac{1}{2d}
\tilde{t}_{(i, p), (j, q)}
&
\{i, j\} \in E
,
\\
0
,
&
\{i, j\} \notin E
,
\end{cases}
\end{align}
where $\block(i,j) = \{(i,p),(j,q)\}$.
We would now like to approximate each $\tilde{t}_{(i, p), (j, q)}$ with 
$t_{(i, p), (j, q)}$, which is the kinetic energy term for a pair of simple Gaussians.
This turns out to be a sufficiently accurate approximation for $\tilde{t}_{(i, p), (i,p)}$.
Note that 
$$\ham[t]{round}_{i, i} = 
\frac12 \left(
t_{\alpha}(0) + t_{\beta}(0)
\right) = \sum_{p}
\psi_{p}^2
t_{(i, p), (i, p)}.$$
However, for edge $\{i,j\}$, where $\block(i,j) = \{(i,p),(j,q)\}$,
primitive orbitals $\phi_{(i,p)}$ and $\phi_{(j,q)}$ are only $\gamma_{i,j}$ apart.
In this case, there is sufficient overlap between the orbitals that the effect of orthonormalizing
the orbitals has a significant impact on the kinetic energy between the pair.
Therefore, instead of setting $\ham[t]{round}_{i, j}$ to be $(1/2d) t_{(i,p),(j,q)}$, we use
a slightly corrected expression as defined by the function $f$. For comparison:
\begin{align}
    \frac{1}{2d} t_{(i,p),(j,q)} & = \frac{\alpha}{4d} (3 - \omega_{i, j}) \exp(-\omega_{i, j}/2),\\
    \ham[t]{round}_{i, j} & = -\frac{\alpha}{4d} \sqrt{f(\omega_{i, j})} = \frac{-\alpha \omega_{i, j}}{4d}  \exp(-\omega_{i, j}/2).
\end{align}

For the potential operator, the coefficients $u$ are a sufficiently good approximation for the
$\tilde{u}$. We will show that we can also drop  potential energy terms that involve
any two primitive orbitals that are a distance at least $\Gamma$ apart. Thus, we only need to include
terms $u_{[(i,p),(j,q)],[(k,r),(l,s)]}$, where the indices $(i,p)$, $(j,q)$, $(k,r)$, and $(l,s)$
are all the same or all come from the set $\block(i,j)$ for some edge $\{i,j\}$.
Thus, $\ham[u]{round}_{i, j, k, l}$ will be $0$, except when $i$, $j$, $k$, $l$ are all equal or are all endpoints
of the same edge.
The rounded potential operator is
\begin{align}
\ham[U]{round} = 
\frac12
\sum_{\substack{(i, j, k, l) \in B
\\
    \sigma \in \pm 1}}
\ham[u]{round}_{i, j, k, l}
\tilde{a}_{i, \sigma}^{\dagger}
\tilde{a}_{j, \tau}^{\dagger}
\tilde{a}_{k, \tau}
\tilde{a}_{l, \sigma}
,
\end{align}
where
\begin{equation}
B = \bigcup_{\{i, j\} \in E} {\{i, j\}}^4
\end{equation}
is the set of all 4-tuples of indices such that they are all the same or there are two distinct indices corresponding to an edge in the graph.
For example, $(i, i, i, i), (i, j, j, i) \in B$  but $(i, i, j, k), (i, k, k, i) \notin B$ for $\{i, j\} \in E$ and $\{i, k\} \notin E$.
The coefficients are defined as
\begin{align}
\ham[c]{round}_U
=
\ham[u]{round}_{i, i, i, i}
&=
\frac14 u_{(i, 0), (i, 0), (i, 0), (i, 0)}
+
\frac1{4d^2}
\sum_{p \in [d]}
u_{(i, p), (i, p), (i, p), (i, p)}
\\
&=
\frac14
\ham[u]{Coul}_{\beta}(0)
+
\frac1{4d}
\ham[u]{Coul}_{\alpha}(0)
,\label{eq:u-round-coul}\\
\ham[u]{round}_{i, j, j, i}
=
\ham[u]{round}_{j, i, i, j}
&=
\frac1{4d^2}
u_{(i, p), (j, q), (j, q), (i, p)}
=
\frac1{4d^2}
\ham[u]{Coul}_{\alpha}(\gamma_{i, j})
,\label{eq:u-round-coul-diff}
\\
\ham[u]{round}_{i, i, j, j}
=
\ham[u]{round}_{j, j, i, i}
=
\ham[u]{round}_{i, j, i, j}
=
\ham[u]{round}_{j, i, j, i}
&=
\frac1{4d^2}
u_{(i, p), (i, p), (j, q), (j, q)}
=
\frac1{4d^2}
\ham[u]{exch}_{\alpha}(\gamma_{i, j})
,\label{eq:u-round-exch}\\
\ham[u]{round}_{i, i, i, j}
=
\ham[u]{round}_{i, i, j, i}
=
\ham[u]{round}_{i, j, i, i}
=
\ham[u]{round}_{j, i, i, i}
&
\\
\ham[u]{round}_{j, j, j, i}
=
\ham[u]{round}_{j, j, i, j}
=
\ham[u]{round}_{j, i, j, j}
=
\ham[u]{round}_{i, j, j, j}
&=
\frac1{4d^2}
u_{(i, p), (i, p), (i, p), (j, q)}
=
\frac1{4d^2}
\ham[u]{other}_{\alpha}(\gamma_{i, j})
,
\end{align}
with $\ham[u]{round}_{i, j, k, l} = 0$ for $(i, j, k, l) \notin B$.
The following lemma bounds the difference between $\ham{ES}$ and $\ham{round}$.
\begin{lemma}[name={},restate={[name=restated]roundinglemma}]\label{lem:RO-error}
If
$\beta \geq \alpha \geq 1$,
$\omega_{\min} \geq 4$,
$\Gamma \ge 640 n^{18} \beta^3$,
and
$\alpha \Gamma^2 \geq 12 \log \beta + 80 \log n + 4 \omega_{\min} + 24$,
then
\begin{equation}
\norm{\ham{ES} - \ham{round}}
\leq
3 n^2 \alpha f(\omega_{\min})
+
\frac{1}{20n^2}  + 8 n^4 \sqrt{\alpha} \cdot \exp(-\omega_{\min}/2)
,
\end{equation}
where
$\omega_{\min} = \alpha \gamma_{\min}^2$. 
\end{lemma}

The matrices $T$, $\tilde{T}$, $S$, and $S^{-1/2}$ are all close to block diagonal. Blocks are either single entries on the diagonal (corresponding to primitive orbitals that are a distance at least $\Gamma$ from all other primitive orbitals) or a $2 \times 2$ sub-matrix corresponding to an edge $\{i,j\}$. Suppose that $\block(i,j) = \{(i, p), (j, q)\}$.
For any $n(d+1) \times n(d+1)$ matrix $A$, let $A_{i,j}$ denote the $2 \times 2$ sub-matrix of $A$ indexed by the elements of $\block(i,j)$:
$$A_{i,j} = 
\begin{pmatrix}
a_{(i,p),(i,p)} & a_{(i,p),(j,q)} \\
a_{(j,q),(i,p)} & a_{(j,q),(j,q)}
\end{pmatrix}.$$
We refer to all of the $A_{i,j}$ blocks collectively as the {\em edge blocks} of $A$. 
The proof of Lemma \ref{lem:RO-error} uses the fact that the off-diagonal terms of $T$ outside of the $T_{i,j}$
blocks are small. The same is true for $\tilde{T}$ and $R=S^{-1/2}$.

$\tilde{U}$ and $U$ are also related by conjugation by $S^{-1/2} \otimes S^{-1/2}$.
We will show that $\tilde{U}$ and $U$ are also close to block diagonal. We define $U_{i,j}$ to be the $4 \times 4$ sub-matrix of $U$ corresponding to the intersections of the four rows and four columns indexed by:
$$[(i,p),(i,p)], ~~[(i,p),(j,q)],~~ [(j,q),(i,p)],~~ [(j,q),(j,q)]$$
The proof of Lemma \ref{lem:RO-error} uses the fact that the off-diagonal terms of $U$ outside of the $U_{i,j}$
blocks are small. The same is true for $\tilde{U}$.
We refer to all of the $U_{i,j}$ blocks collectively as the {\em edge blocks}.

\cref{lem:RO-error} is proved in~\cref{sec:rounding-error}.
Outside of this subsection, all creation and annihilation operators are those of the orthonormalized basis $\tilde{\boldsymbol \phi}$; in other words, we'll drop the tildes.

\subsection{Getting the main Hamiltonian}\label{sec:approx}
With the rounded Hamiltonian $\ham{round}$ in hand, we make one final approximation to get to the main Hamiltonian $\ham{main}$ that we will later show is close to a Hubbard Hamiltonian.
Specifically, we remove the ``off-diagonal'' Coulomb interaction terms.
The error of this approximation is bounded by~\cref{lem:offsite-bound}.

The main Hamiltonian is
\begin{align}
\ham{round} 
&= \ham{main} + \ham{approx} + n \cdot c_T,
\\
\ham{main} 
&=
\ham[c]{main}_U
\sum_{i} n_{i, +1} n_{i, -1}
+
\sum_{\substack{
\{i, j\} \in E \\
\sigma \in \{\pm 1\}
}}
\ham[t]{round}_{i, j}
\left(
a_{i, \sigma}^{\dagger} a_{j, \sigma}
+
a_{j, \sigma}^{\dagger} a_{i, \sigma}
\right) 
,
\end{align}
where
$
\ham[c]{main}_U
=
\ham[u]{Coul}_{\beta}(0) / 4
$
.
The difference $\ham{round} - \ham{main} - n \cdot c_T$ contains two types of terms, both of whose coefficients are $O(\sqrt{\alpha})$: the smaller part of the onsite terms $\ham[c]{round}_U$, and the offsite terms corresponding to edges in the interaction graph.
The following lemma bounds the contribution from this difference..
\begin{lemma}\label{lem:offsite-bound}
\begin{align}
\norm{\ham{round} - \ham{main} - n c_T} \leq 
30 n^2 \sqrt{\alpha}.
\end{align}
\end{lemma}

\begin{proof}[Proof of \cref{lem:offsite-bound}]
First, recall that we're restricting to the fixed-particle number subspace, in which the diagonal part $c_T \sum_{i, \sigma} n_{i, \sigma}$ of $\ham[T]{round}$ is the constant $n \cdot c_T$.
That is, $\ham[T]{round} = \ham[T]{main} + n \cdot c_T$.
Let 
\begin{equation}
B_2 = B \setminus \left\{(i, i, i, i) : i \in [n] \right\}
\end{equation}
be the subset of $B$ whose elements contain two distinct indices (corresponding to an edge).
\begin{align}
\norm{\ham{round} - \ham{main} - n \cdot c_T}
&=
\norm{\ham[U]{round} - \ham[U]{main}}
\\
&=
\norm{
\frac12
\sum_{\substack{(i, j, k, l) \in B \\ \sigma, \tau \in \{\pm 1\}}}
\ham[u]{round}_{i, j, k, l}
a_{i, \sigma}^{\dagger}
a_{j, \tau}^{\dagger}
a_{k, \tau}
a_{l, \sigma}
-
\ham[c]{main}_U
\sum_{i \in [n]}
n_{i, +1} n_{i, -1}
}
\\
&\leq
\sum_{i \in [n]}
\left|
\ham[c]{round}_{U}
-
\ham[c]{main}_{U}
\right|
+
\frac12
\sum_{\substack{(i, j, k, l) \in B_2 \\
\sigma, \tau \in \{\pm 1\}
}}
\ham[u]{round}_{i, j, k, l}
\\
&=
n \cdot \frac{1}{4d} \ham[u]{Coul}_{\alpha}(0)
+
\frac12
\sum_{\substack{(i, j, k, l) \in B_2 \\
\sigma, \tau \in \{\pm 1\}
}}
\ham[u]{round}_{i, j, k, l}
\\
&\leq
n \cdot \frac{1}{4d} \ham[u]{Coul}_{\alpha}(0)
+
\frac12
\cdot
\underbrace{4}_{\sigma, \tau}
\cdot
\underbrace{14 \cdot \binom{n}{2} }_{B_2}
\cdot
\ham[u]{Coul}_{\alpha}(0)
\\
&\leq
15 n^2 \ham[u]{Coul}_{\alpha}(0)
=
15 n^2 \cdot
    \frac{2}{\sqrt{\pi}} \sqrt{\alpha}
\leq 30 n^2 \sqrt{\alpha}.
\end{align}
\end{proof}

\subsection{Hardness of estimating ground state energy}\label{sec:qma-proof}

Now, we're ready to prove the main theorem.

\begin{proof}[Proof of \cref{thm:electronic-structure}]
Membership in QMA is straightforward.
For hardness, we reduce from the Fermi-Hubbard model.
Recall~\cref{thm:hubbard}: for some $p$, $q$ and all 
$\ham[u]{Hubb}_{0} \geq n^{14 + 3p + 2q}$,
finding the ground state to precision $n^{-q}$ of 
\begin{equation}
\ham{Hubb}
=
\ham[u]{Hubb}_0 \sum_{i \in [n]}
n_{i, +1} n_{i, -1}
+
\sum_{i < j}
\ham[t]{Hubb}_{i, j}
\left(
a_{i, \sigma}^{\dagger}
a_{j, \sigma}
+
a_{j, \sigma}^{\dagger}
a_{i, \sigma}
\right)
\end{equation}
subject to $\left|\ham[t]{Hubb}_{i, j}\right| \leq \sqrt{n^p \ham[u]{Hubb}_0}$
is QMA-complete.
In the preceding sections, we showed that, using our choice of single-electron orbitals, the electronic structure Hamiltonian is close to
\begin{equation}
\ham{ES}
\approx
\ham{main}
+ 
n \cdot c_T
=
\ham[c]{main}_U
n_{i, +1} n_{i, -1}
+
\sum_{i < j}
\ham[t]{round}_{i, j}
\left(
a_{i, \sigma}^{\dagger}
a_{j, \sigma}
+
a_{j, \sigma}^{\dagger}
a_{i, \sigma}
\right)
+
n \cdot c_T
.
\end{equation}

To prove the theorem, it suffices to show that for any Hubbard Hamiltonian satisfying the conditions of~\cref{thm:hubbard}, 
we can set the parameters $\alpha, \beta, {\left\{\gamma_{i, j}\right\}}_{i, j}, \Gamma$ such that
\begin{equation}
\rho \ham{Hubb} = \ham{main}
\end{equation}
and
\begin{equation}
\norm{\ham{main} - \ham{ES} - n \cdot c_T}
=
o\left(\rho n^{-q}\right)
\label{eq:main-ES-error}
\end{equation}
for some $\rho \in \mathbb R$.
With this, finding the ground state of $\ham{ES}$ to precision $O(\rho n^{-q})$ would allow us to find the ground state of $\ham{Hubb}$ to precision $O(n^{-q})$, and so the former must be QMA-hard.
We'll base our parameterization on four constants independent of $n$:
\begin{align}
a &= \log_n \alpha, &
b &= \log_n \beta, &
r &= \log_n \rho, &
g &= -\log_n \sqrt{f(\omega_{0})} = -\frac12 \log_n f(\omega_{0}),
\end{align}
where $\omega_0$ is a lower bound on $\omega_{i, j}$ to be set later.
The first three immediately set $\alpha$, $\beta$, and $\rho$, respectively.
Equating $n^r \ham{Hubb}$ and $\ham{main}$ requires
\begin{align}
n^r \ham[u]{Hubb}_0 &= \ham[c]{main}_U = 
\frac{1}{2\sqrt{\pi}} \sqrt{\beta}
,
\\
n^r \ham[t]{Hubb}_{i, j}
&=
-\frac{\alpha}{4d} \sqrt{f(\omega_{i, j})}
.
\label{eq:coeff-eq}
\end{align}

\paragraph{Coefficient ranges}

If we set
\begin{equation}
\boxed{
b = 30 + 6p + 4q + 2r
}
\end{equation}
then
\begin{align}
\ham[u]{Hubb}_0
&= 
\frac{\sqrt{\beta}}{2\sqrt{\pi} n^r}
\geq
\frac14
\frac{\sqrt{\beta}}{n^r}
=
4^{-1}
n^{0.5 b - r}
=
\frac{n}{4}
n^{14 + 3 p + 2 q}
\geq
n^{14 + 3 p + 2 q}
&
n \geq 4
\end{align}
satisfies the lower bound in the statement of~\cref{thm:hubbard}.

If we set
\begin{equation}
\boxed{
g = -\frac12 p + a - \frac14 b - \frac32 - \frac12 r
\geq 1
}
,
\label{eq:g-eq}
\end{equation}
then for $n \geq 9$,
\begin{align}
\frac{\alpha}{4d} \sqrt{f(\omega_0)}
&
\geq
\frac{\alpha}{4n} \sqrt{f(\omega_0)}
=
\frac14 n^{a - g - 1}
=
\frac{\sqrt{n}}{4} n^{\frac12 r + \frac12 p + \frac14 b}
\\
&\geq
\frac{3}{4} n^{\frac12 r + \frac12 p + \frac14 b}
\\
&
\geq
\frac{1}{\sqrt{2\sqrt{\pi}}}
n^{\frac12 r + \frac12 p + \frac14 b}
=
n^r
\sqrt{
n^p
\frac{n^{-r}\sqrt{\beta}}{2\sqrt{\pi}}
}
=
\rho
\sqrt{n^p \ham[u]{Hubb}_0}
,
\end{align}
and thus for any $\ham[t]{Hubb}_{i, j} \leq \sqrt{n^p \ham[u]{Hubb}_0}$ there is some $\omega_{i, j} \geq \omega_0$ that satisfies~\cref{eq:coeff-eq}.

\paragraph{Bounding the difference between ES and Hubbard}
The difference between the electronic structure Hamiltonian and the main Hamiltonian is
\begin{align}
\norm{\ham{main} + n \cdot c_T- \ham{ES}}
&\leq 
\norm{\ham{main} + n \cdot c_T - \ham{round}}
+
\norm{\ham{round} - \ham{ES}}
\\
&\leq
30 n^2 \sqrt{\alpha} + 
\frac1{20n^2}
+ 3 n^2 \alpha f(\omega_{\min}) + 8 n^4 \sqrt{\alpha} \exp(-\omega_{\min} / 2)
\\
&\leq
30 n^2 \sqrt{\alpha} 
+
\frac1{20n^2}
+ 3 n^2 \alpha f(\omega_{0}) + 8 n^4 \sqrt{\alpha} \sqrt{f(\omega_{0})}
\\
&=
O\left(
n^{2 + \frac12 a} + 
n^{-2}
+
n^{2 + a -2g}
+
n^{4 + \frac12 a - g}
\right)
\\
&
=
O\left(
n^{4 + \frac12 a} + 
n^{2 + a -2g}
\right)
.
\end{align}
Therefore, to satisfy~\cref{eq:main-ES-error}, it would suffice to have
\begin{align}
\boxed{
4 + \frac12 a
< r - q
}
\end{align}
and
\begin{align}
2 + a - 2g
< r - q
.
\end{align}
Plugging~\cref{eq:g-eq} into the latter yields
\begin{equation}
\boxed{
p
+q
+5
< a - \frac12 b
}
.
\end{equation}

\paragraph{Parameter setting}
In summary, our constraints are
\begin{align}
30 + 6 p + 4 q &= b - 2r, \\
\frac12 p + \frac52 &< a - \frac14 b - \frac12 r,
\\
4 + q & < -\frac12a + r,
\\
p + q + 5 & < a - \frac12 b.
\end{align}
The following settings satisfy all the required constraints:
\begin{align}
a & = 18p + 12q + 90,
\\
b &= \frac53 a,
\\
r &= \frac23 a,
\\
g &= \frac14 a - \frac12 p - \frac32.
\end{align}

\end{proof}

\section{Hardness of finding lowest-energy Slater determinant}
\label{sec:hf}

In this section, we show that finding the lowest-energy Slater determinant (i.e., Hartree-Fock state) of an electronic structure Hamiltonian is NP-hard.
This is a natural complement to our QMA-hardness result, in that Slater determinants are the most natural class of fermionic states that are efficiently representable and manipulable classically.
The proof has much in common with that of~\cref{thm:electronic-structure}.
We start with the same parameterized construction of orbitals described in~\cref{sec:orbitals}, and then orthonormalize and round them as in~\cref{sec:nonorthonormality} to get the Hamiltonian $\ham{round}$.
We then diverge from the QMA-hardness proof by setting the parameters in a different regime.
Specifically, we set the exponents $\alpha$ and $\beta$ large enough that the Hamiltonian becomes essentially classical (diagonal).
The proof concludes by showing that this classical Hamiltonian can express an NP-hard problem such as independent set.

\begin{customthm}{\labelcref*{thm:LESD-NP-hard-informal}}\label{thm:ES-SD-NP-hard}
Determining the lowest-energy Slater determinant of an electronic structure Hamiltonian in a fixed basis and with fixed particle number to inverse-polynomial precision is NP-complete.
\end{customthm}

\begin{proof}
To start, we'll set $\gamma_{i, j} = \gamma$ for all $\{i, j\} \in E$.
We show that the parameters $\alpha$, $\beta$, $\gamma$, $\Gamma$ can be set such that the electronic structure approximates a diagonal Hamiltonian
\begin{equation}
\ham{ES} 
-
n \cdot c_T
\approx
\ham{class}
=
\ham[u]{class}_1
\sum_i 
n_{i, +1} n_{i, -1}
+
\ham[u]{class}_2
\sum_{\substack{\{i, j\} \in E \\ \sigma, \tau \in \{\pm 1\}}} 
n_{i, \sigma} n_{j, \tau},
\end{equation}
where
\begin{align}
\ham[u]{class}_1 &= \ham[c]{round}_U
,
&
\ham[u]{class}_2 &= \frac1{4d^2} \ham[u]{Coul}_{\alpha}(\gamma).
\end{align}
For a diagonal Hamiltonian, there is always a computational basis state of lowest energy.
Because basis states are a special case of Slater-determinants, finding the lowest-energy Slater-determinant for diagonal Hamiltonians is equivalent to finding the ground state.

For sufficiently large $\ham[u]{class}_1 > 4 n^2 \ham[u]{class}_2$, the ground space of $\ham{class}$ in the $k$-electron subspace for $k \leq n$  will have at most one electron in each spatial orbital, and the ground state energy is
\begin{align}
h(\mathbf n)
=
\ham[u]{class}_2
\sum_{\{i, j\} \in E}
n_i n_j,
\end{align}
where $n_i = n_{i, +1} + n_{i, -1}$ is the occupancy of the $i$-th spatial orbital.

The state space is spanned by vectors $\mathbf n$ such that $\sum_i n_i = k$, which we can interpret as representing a subset $S \subset V$ of vertices with size $|S| = k$.
The classical function $h(\mathbf n)$ is then proportional to the number of edges with both endpoints in the set $S$.
In other words, if $h(\mathbf n) = 0$, then the set $S$ is an independent set of size $k$; otherwise $h(\mathbf n) \geq \ham[u]{class}_2$.
Therefore, if $\ham[u]{class}_1$ is sufficiently larger than $\ham[u]{class}_2$, then 
finding the lowest-energy Slater-determinant of $\ham{class}$ in the $k$-electron subspace to precision $\ham[u]{class}_2$ is as hard as determining if a graph has an independent set of size $k$.

To finish the proof, we just need to set the parameters such that
\begin{align}
\ham[u]{class}_1 & > 4 n^2 \ham[u]{class}_2,
\\
\norm{\ham{ES} - \ham{class}} & < \frac12 \ham[u]{class}_2.
\end{align}

Let $\gamma = 1$, leaving $\alpha$, $\beta$, and $\Gamma$ to be set.
The first constraint is satisfied by $\beta \geq 16 n^4$:
\begin{align}
\\
\ham[u]{class}_1 &= \ham[c]{round}_U \geq \frac14 \ham[u]{Coul}_{\beta}(0)
\\
&= \frac14 \sqrt{\frac{4 \beta}{\pi}} > \frac14 \sqrt{\beta}
\\
& \geq 
n^2 
\\
&\geq
\frac{n^2}{d^2} \erf\left(\sqrt{\alpha \gamma^2}\right)
\\
&=
4n^2 \ham[u]{class}_2
.
\end{align}
For the second constraint, if $\alpha \geq 1$, then
\begin{equation}
\frac12 \ham[u]{class}_2
\geq
\frac12 \frac{1}{4d^2} \erf{1}
\geq
\frac{1}{8n^2} \cdot \frac12 \geq \frac{1}{16 n^2}.
\end{equation}

\begin{lemma}\label{lem:round-class-diff}
For $\alpha \geq 1$, $\gamma_{i, j} = \gamma \geq 1$,
\begin{equation}
\norm{
\ham{round}
-
\ham{class}
}
\leq
14 \alpha n^2  e^{-\alpha \gamma^2 / 4}.
\end{equation}
\end{lemma}

\begin{proof}[Proof of~\cref{lem:round-class-diff}]
The classical Hamiltonian $\ham{class}$ has no kinetic component, and so we need to bound the entirety of the non-constant kinetic component of the rounded Hamiltonian $\ham{round}$ :
\begin{align}
\norm{
\ham[T]{round}
-
n\cdot c_T
}
&=
\norm{
\sum_{\substack{\{i, j\} \in E \\ \sigma \in \{\pm 1\}}} \ham[t]{round}_{i, j}}
\left(
a_{i, \sigma}^{\dagger} a_{j, \sigma}
+
a_{j, \sigma}^{\dagger} a_{i, \sigma}
\right)
\\
&\leq
4 \sum_{\{i, j\} \in E} \left|\ham[t]{round}_{i, j}\right|
\\
&\leq
\frac{\alpha}{d} n^2 \sqrt{f(\omega)}
\leq
\alpha^2 n^2 \gamma^2 e^{-\alpha \gamma^2 /2} 
.
\end{align}
For the potential difference, define
\begin{equation}
B_3 = B_2 \setminus \left\{(i, j, j, i) : \{i, j\} \in E\right\},
\end{equation}
i.e. the indices of potential terms that are \emph{not} Coulomb (which are exactly those included in $\ham{class}$).
Then
\begin{align}
&\norm{
\ham[U]{round}
-
\ham{class}
}
\\
&=
\norm{
\frac12
\sum_{\substack{(i, j, k, l) \in B \\ \sigma, \tau \in \{\pm 1\}}}
\ham[u]{round}_{i, j, k, l}
a_{i, \sigma}^{\dagger}
a_{j, \tau}^{\dagger}
a_{k, \tau}
a_{l, \sigma}
-
\ham[u]{class}_1
\sum_{i \in [n]}
n_{i, +1} n_{i, -1}
-
\ham[u]{class}_2
\sum_{\substack{\{i, j\} \in E \\ \sigma, \tau \in \{\pm 1\}}} 
n_{i, \sigma} n_{j, \tau},
}
\\
&\leq
\frac12
\sum_{\substack{(i, j, k, l) \in B_3 \\
\sigma, \tau \in \{\pm 1\}
}}
\ham[u]{round}_{i, j, k, l}
\\
&\leq
\frac12
\cdot
\underbrace{4}_{\sigma, \tau}
\cdot
\underbrace{12 \cdot \binom{n}{2}}_{B_3}
\cdot
\frac1{4d^2}
2 \sqrt{\alpha}
\exp\left(-\alpha \gamma^2 / 2\right)
\\
&\leq
6 \sqrt{\alpha} n^2 
\exp\left(-\alpha \gamma^2 / 2\right)
.
\end{align}
Putting them together,
\begin{align}
\norm{
\ham{round}
-
\ham{class}
}
&\leq
\norm{
\ham[T]{round}
-
n\cdot c_T
}
+
\norm{
\ham[U]{round}
-
\ham{class}
-
n\cdot c_T
}
\\
&\leq 
7 \alpha n^2 (\alpha \gamma^2) e^{-\alpha \gamma^2 / 2}
\leq
14 \alpha n^2  e^{-\alpha \gamma^2 / 4}
\end{align}
\end{proof}

Together, \cref{lem:RO-error,lem:round-class-diff} imply that for
$\gamma = 1$,
$\beta \geq \alpha > 74 + 48 \log n$,
$\Gamma \ge 640 n^{18} \beta^3$,
and
$\alpha \Gamma^2 \geq 12 \log \beta + 80 \log n + 4 \alpha + 24$,
\begin{align}
\norm{
\ham{ES}
-
\ham{class}
-
n\cdot c_T
}
&\leq
\norm{
\ham{ES}
-
\ham{round}
}
+
\norm{
\ham{round}
-
\ham{class}
-
n\cdot c_T
}
\\
&\leq 
3 n^2 \alpha f(\omega_{\min})
+
\frac{1}{20n^2}
+ 
8 n^4 \sqrt{\alpha} \cdot \exp(-\omega_{\min}/2)
+
14 \alpha n^2 e^{-\alpha \gamma^2 / 4}
\\
&=
3 n^2 \alpha^2 e^{-\alpha /2}
+
\frac{1}{20n^2}
+ 
8 n^4 \sqrt{\alpha} \cdot \exp(-\alpha/2)
+
14 \alpha n^2 e^{-\alpha / 4}
\\
&\leq 
\frac1{20n^2}
+
100 n^4 e^{-\alpha / 8}
\leq
\frac1{20n^2}
+
\frac1{80n^2}
=
\frac1{16n^2}
\end{align}
where used the fact that for $x \geq 0$, $\max\{x^2 e^{-x/2}, \sqrt{x} e^{-x/2}, x e^{-x/4}\} \leq 4 e^{-x/8}$.
For sufficiently large $n$, it suffices to set $\beta = \alpha = n$, $\gamma =1$, and $\Gamma = n^{32}$.
\end{proof}

\section{Acknowledgements}

We are grateful to the Simons Institute for the Theory of Computing, at whose program on the ``The Quantum Wave in Computing'' this collaboration began.
B.O. is supported by a NASA Space Technology Research Fellowship. B.F. acknowledges support from AFOSR (YIP number FA9550-18-1-0148 and FA9550-21-1-0008). This material is based upon work partially supported by the National Science Foundation under Grant CCF-2044923 (CAREER). J.D.W. is supported by funded by the NSF (PHYS-1820747) and the Department of Energy (Grant DE-SC0019374). JDW is also supported by NSF (EPSCoR-1921199) and by the U.S. Department of Energy, Office of Science, Office of Advanced Scientific Computing Research under programs Quantum Computing Application Teams and Accelerated Research for Quantum Computing program.

\appendix

\section{Proof of Lemma ~\ref{lem:RO-error}}
\label{sec:rounding-error}

The proof of Lemma \ref{lem:RO-error} uses the following technical lemmas that quantify the statement that  matrices $\tilde{T}$,
$S$, and $\tilde{U}$
are approximately block diagonal.
It will be convenient to refer only to the entries along the diagonal or inside the edge blocks. For an $n(d+1) \times n(d+1)$ matrix or an $n^2(d+1)^2 \times n^2(d+1)^2$ matrix $A$, let $\ham[A]{block}$ to denote the matrix obtained by replacing all of the off-diagonal entries of $A$ outside the edge blocks with $0$. 
Define 
\begin{align}
    \ham[S]{neg} & = S - \ham[S]{block},\\ 
    R & = S^{-1/2},\\
    \ham[R]{aprx} &= (\ham[S]{block})^{-1/2},\\
    \ham[R]{neg} & = R - \ham[R]{aprx}.
\end{align}
Note that, because $\ham[S]{block}$ is block diagonal, $\ham[R]{aprx}$ is also block diagonal. However $\ham[R]{block} \neq \ham[R]{aprx}$.
The matrix $\ham[R]{neg}$, unlike $\ham[S]{neg}$, has non-zero entries even on the diagonal and within the blocks, though these are small.

The first lemma bounds $\max(|\ham[R]{neg}|)$,
where $\max(|A|)$ is defined to be
the maximum of the absolute values of the entries
in matrix $A$.

\begin{lemma}\label{lem:r-bound}
If $\alpha \Gamma^2  \geq  4 \log n + 2\omega_{\min} + 2$ and $\omega_{\min} \geq 4$, then
\begin{align}
\ham[r]{neg}_{\max}
=
\max\left( \left| \ham[R]{neg} \right| \right)
& 
\le n^2 \exp\left[-(\alpha \Gamma^2-\omega_{\min})/2\right]
.
\label{eq:r-neg-bound}
\end{align}
\end{lemma}
\begin{corollary}\label{cor:r-neg}
For $\alpha \Gamma^2 \geq 4\log n + 2\omega_{\min} + 2$ and $\omega_{\min} \geq 2$, 
\begin{equation}
\ham[r]{neg}_{\max} \leq 
n^2 \exp\left[-\left(
\alpha \Gamma^2
-
\omega_{\min} 
\right)/2\right]
\leq
n^2 \exp\left[-\left(
4 \log n + 2\omega_{\min} + 2
-
\omega_{\min} 
\right)/2\right]
\leq
\exp(-1)
\leq
1/2.
\label{eq:r-neg-bound-constant}
\end{equation}
\end{corollary}

The entries of matrix $S$ are just the overlap of normalized Gaussians,
so the diagonal is all ones. Block $S_{i,j}$ corresponding to edge 
$\{i,j\}$, where $\block(i,j) = \{ (i,p), (j,q)\}$ is
\begin{align}
S_{i, j} 
&=
\begin{pmatrix}
1 & \epsilon_{i, j} \\
\epsilon_{i, j} & 1
\end{pmatrix}
,
\\
\epsilon_{i, j} &= s_{(i, p), (j, q)}
=
s_{\alpha}(\gamma_{i, j}) = \exp(-\omega_{i,j}/2),
    & \text{where }\omega_{i,j} = \alpha \gamma_{i,j}^2
.
\end{align}

The entry $s_{(i,p),(i,p)}$ is not contained in an edge block if and only if
$p=0$ or $p > \degree_i$. In this case,
the orbital $\phi_{i,p}$ is at least a distance $\Gamma$ away from every other
primitive orbital, and the block for $s_{(i,p),(i,p)}$ is
just the single element on the diagonal. For these primitive orbitals, we have
$$s_{(i,p),(i,p)} = \ham[r]{aprx}_{(i,p),(i,p)} = 1.$$
The edge blocks of $\ham[R]{aprx}$ can be computed exactly as
\begin{align}
\ham[R]{aprx}_{i, j}
&=
{\left(S_{i, j}\right)}^{-1/2}
\\
&=
\frac12
\begin{pmatrix}
\frac1{\sqrt{1+\epsilon_{i, j}}}
+
\frac1{\sqrt{1-\epsilon_{i, j}}}
&
\frac1{\sqrt{1+\epsilon_{i, j}}}
-
\frac1{\sqrt{1-\epsilon_{i, j}}}
\\
\frac1{\sqrt{1+\epsilon_{i, j}}}
-
\frac1{\sqrt{1-\epsilon_{i, j}}}
&
\frac1{\sqrt{1+\epsilon_{i, j}}}
+
\frac1{\sqrt{1-\epsilon_{i, j}}}
\end{pmatrix}
.
\end{align}
The following lemma bounds the error from just taking the leading term in $\epsilon_{i, j}$. Note that the $2 \times 2$ matrix $\ham[R]{aprx}_{i,j}$
has identical on-diagonal entries and identical off-diagonal entries.
Let $\on(\ham[R]{aprx}_{i,j})$ refer to the value of the on-diagonal
entries and let $\off(\ham[R]{aprx}_{i,j})$ refer to the value of the off-diagonal
entries. The matrix $\ham[R]{aprx}_{i,j} T_{i,j} \ham[R]{aprx}_{i,j}$
has the same symmetries, so we can define $\on$ and $\off$ for those matrices as well.
\begin{lemma}\label{lem:r-approx}
For $\omega_{\min} \geq 4$ and $\{i, j\} \in E$ where
$\block(i,j) = \{(i,p),(j,q)\}$, 
\begin{align}
1 & \leq 
\on(\ham[r]{aprx})
\leq 
1 + \epsilon_{i, j}^2
,
\label{eq:r-block-diag-bounds}
\\
-\frac{\epsilon_{i, j}}2 - \epsilon_{i, j}^3
&\leq
\off(\ham[r]{aprx})
\leq
-\frac{\epsilon_{i, j}}2
,
\label{eq:r-block-off-diag-bounds}
\\
t_{\alpha}(0)
&\leq
{\on\left(
\ham[R]{aprx}_{i, j}
T_{i, j}
\ham[R]{aprx}_{i, j}
\right)}
\leq
t_{\alpha}(0)
+
\alpha \omega_{i, j} \epsilon_{i, j}^2
,
\label{eq:RTR-diag-bounds}
\\
-\frac{\alpha}{2}\sqrt{f(\omega_{i, j})}
(1 + 4\epsilon_{i, j}^2)
&
\leq
{\off\left(
\ham[R]{aprx}_{i, j}
T_{i, j}
\ham[R]{aprx}_{i, j}
\right)}
\leq
-\frac{\alpha}{2}\sqrt{f(\omega_{i, j})}
,
\label{eq:RTR-off-diag-bounds}
\end{align}
\begin{align}
\max \left(
\left|
{
{\left(
\ham[R]{aprx}_{i, j}
\right)}^{\otimes 2}
U_{i, j}
{\left(
\ham[R]{aprx}_{i, j}
\right)}^{\otimes 2}
-
U_{i, j}
}
\right|\right)
&\leq
16 \sqrt{\alpha} \epsilon_{i, j}
.
\label{eq:RRURR-bound}
\end{align}
\end{lemma}

\begin{corollary}\label{cor:r-max}
For $\omega_{\min} \geq 4$, 
\begin{align}
\ham[r]{aprx}_{\max}
&=
\max \left(
\left|
\ham[R]{aprx}
\right| \right)
\leq 3/ 2
.
\label{eq:r-block-bound}
\end{align}
\end{corollary}
\begin{corollary}
For $\alpha \Gamma^2 \geq 4\log n + \omega_{\min} + 2$ and $\omega_{\min} \geq 4$, 
\begin{align}
r_{\max} 
&= \max \left(
\left|
R
\right| \right)
\leq
\ham[r]{aprx}_{\max} + \ham[r]{neg}_{\max}
\leq 2
.
\label{eq:r-bound}
\end{align}
\end{corollary}

Define 
$\ham[T]{neg} = T - \ham[T]{block}$.
Similarly, define
$\ham[U]{neg} = U - \ham[U]{block}$.
The following lemma bounds these coefficients.
\begin{lemma}\label{lem:integral-bounds}
For $\beta \geq \alpha \geq 1$ and $\alpha \Gamma^2 \geq 64$,
\begin{align}
t_{\max} 
&=
\max\left(
\left|
T
\right| \right)
\leq
\frac32 \beta
,
\label{eq:t-bound}
\\
\ham[t]{neg}_{\max}
&=
\max\left(
\left|
\ham[T]{neg}
\right| \right)
\leq 
\beta
\exp\left(-\alpha \Gamma^2 / 4\right),
\label{eq:t-neg-bound}
\\
u_{\max}
&=
\max\left(
\left|
U
\right| \right)
\leq
2\beta^3
,
\label{eq:u-bound}
\\
\ham[u]{neg}_{\max}
&=
\max\left(
\left|
\ham[U]{neg}
\right| \right)
\leq 2 \beta^3 / \Gamma
.
\label{eq:u-neg-bound}
\end{align}
\end{lemma}

Proofs of the technical lemmas follow the proof of~\cref{lem:RO-error}.
Note that the conditions of the technical lemmas (and corollaries) are implied by the conditions of~\cref{lem:RO-error}.

\roundinglemma*

\begin{proof}[Proof of Lemma ~\ref{lem:RO-error}]

We will bound the kinetic and potential parts separately, starting with the former.

Define $\ham[\tilde{T}]{aprx} = \ham[R]{aprx} \ham[T]{block} \ham[R]{aprx}$,
and recall that $\tilde{T} = RTR$.
The first task is to bound the error of approximating
$\tilde{T}$ by $\ham[\tilde{T}]{aprx}$:
\begin{align}
    \max\left( \left| \tilde{T} - \ham[\tilde{T}]{aprx}\right| \right) 
&= \max\left( \left| RTR - \ham[R]{aprx} \ham[T]{block} \ham[R]{aprx}\right| \right)\\
& \le \max\left( \left| RTR - R \ham[T]{block} R\right| \right)  + \max\left( \left| R\ham[T]{block} R - \ham[R]{aprx} \ham[T]{block} \ham[R]{aprx}\right| \right)
.
\label{eq:TtildeBound}
\end{align} 
We will bound each term from (\ref{eq:TtildeBound}) separately.
We will use the fact that if $A$ and $B$ are $m \times m$ matrices,
then $\max(|AB|) \le m \cdot \max(|A|) \cdot \max(|B|)$.
Since the matrices $R$ and $T$ are $n(d+1) \times n(d+1)$ matrices
and $d+1 \le n$, we will pick up a factor of at most $n^2$ every time
this rule is applied. 
\begin{align} 
\max\left( \left| RTR - R \ham[T]{block} R\right| \right) & =
\max\left( | R (T-\ham[T]{block}) R |\right)\\
& \le n^4 (r_{\max})^2 \max(| T-\ham[T]{block}|) = n^4 (r_{\max})^2  \ham[t]{neg}_{\max}\\
& \le 4 n^2 \beta \exp(-\alpha \Gamma^2/4).
 \label{eq:tTilde-bound-1}
\end{align}
The last inequality uses the bound from (\ref{eq:r-bound}) that $r_{\max} \le 2$
and from (\ref{eq:t-neg-bound}) that $\ham[t]{neg}_{\max} \le \beta \exp(-\alpha \Gamma^2/4)$.
To bound the second term from (\ref{eq:TtildeBound}),
recall that $R = \ham[R]{aprx} + \ham[R]{neg}$.
\begin{align}
& \max\left( \left| R\ham[T]{block} R - \ham[R]{aprx} \ham[T]{block} \ham[R]{aprx}\right| \right)\\
& = \max\left( \left| \left[ \ham[R]{aprx} + \ham[R]{neg} \right] \ham[T]{block} \left[ \ham[R]{aprx} + \ham[R]{neg} \right] - \ham[R]{aprx} \ham[T]{block} \ham[R]{aprx}\right| \right)\\
& =  \max\left( \left|  \ham[R]{aprx}  \ham[T]{block}  \ham[R]{neg} + 
\ham[R]{neg}  \ham[T]{block}  \ham[R]{aprx} + \ham[R]{neg}  \ham[T]{block}  \ham[R]{neg}\right| \right)\\
& \le n^4 t_{\max} \ham[r]{neg}_{\max} \left( 2 \ham[r]{aprx}_{\max}+  \ham[r]{neg}_{\max} \right)\\
& \le n^4 \underbrace{\frac 3 2 \beta}_{(\ref{eq:t-bound})}
\underbrace{ \exp\left[-(\alpha \Gamma^2-\omega_{\min})/2\right]}_{(\ref{eq:r-neg-bound})}
 \left( 2 \underbrace{\frac 3 2}_{(\ref{eq:r-block-bound})} +  \underbrace{\frac 1 2}_{(\ref{eq:r-neg-bound-constant})} \right)\\
 & \le 6 n^4 \beta \exp\left[-(\alpha \Gamma^2-\omega_{\min})/2\right] \le 6 n^4 \beta \exp(-\alpha \Gamma^2/4)
 \label{eq:tTilde-bound-2}
\end{align}
The last inequality is implied by the assumptions of the lemma, specifically
that $\alpha \Gamma^2 \ge 2 \omega_{\min}$.
Putting together the bounds from (\ref{eq:tTilde-bound-1}) and (\ref{eq:tTilde-bound-2})
we get that
\begin{equation}
 \max\left( \left| \tilde{T} - \ham[\tilde{T}]{aprx}\right| \right)  \le 10 n^4 \beta 
 \exp(-\alpha \Gamma^2/4).
 \label{eq:t-approx-bound}
\end{equation}
The next step is to use the approximation for the kinetic-energy terms
for the primitive orbitals to get the kinetic-energy term for the composite orbitals.
Recall that composite orbital $\phi_i$ is a superposition of $\phi_{i,p}$:
$$\phi_i = \sum_{p=0}^d \psi_p \phi_{i,p},$$
where $\psi_0 = 1/\sqrt{2}$ and $\psi_{p>0} = 1/\sqrt{2d}$.
Therefore, the kinetic-energy terms for the composite orbitals are just
superpositions of the kinetic-energy terms for the primitive  orbitals:
$$t_{i,j} = \sum_{p,q} \psi_p \psi_q t_{(i,p),(j,q)}.$$
We can apply this principle to $\tilde{T}$ and $\ham[\tilde{T}]{aprx}$ as well:
$$\tilde{t}_{i,j} = \sum_{p,q} \psi_p \psi_q \tilde{t}_{(i,p),(j,q)}
~~~~~\mbox{and}~~~~~
\ham[\tilde{t}]{aprx}_{i,j} = \sum_{p,q} \psi_p \psi_q \ham[\tilde{t}]{aprx}_{(i,p),(j,q)}.$$
Using the bound from (\ref{eq:t-approx-bound}):
\begin{align}
    |\tilde{t}_{i,j} - \ham[\tilde{t}]{aprx}_{i,j}|
    & \le \sum_{p,q} \psi_p \psi_q | \tilde{t}_{(i,p),(j,q)} - \ham[\tilde{t}]{aprx}_{(i,p),(j,q)}| \\
    & \le \frac 1 2 (d+1)^2 \max\left( \left| \tilde{T} - \ham[\tilde{T}]{aprx}\right| \right)
    \le 5 n^6 \beta  \exp(-\alpha \Gamma^2/4).
    \label{eq:tilde-aprx}
\end{align}

The next task is to bound $|\ham[\tilde{t}]{aprx}_{i,j} - \ham[t]{round}_{i,j}|$.
We will consider three separate cases. In each case, we will show that 
\begin{equation}
\left|\ham[\tilde{t}]{aprx}_{i, j} - \ham[t]{round}_{i, j}\right| \leq \alpha f(\omega_{\min})
.
\label{eq:t-aprx-t-round-diff}
\end{equation}
Recall that $\ham[\tilde{T}]{aprx} = \ham[R]{aprx} \ham[T]{block} \ham[R]{aprx}$,
so  matrix $\ham[\tilde{T}]{aprx}$ is block diagonal. This means
that $\ham[\tilde{t}]{aprx}_{(i,p),(j,q)} = 0$ unless
$(i,p) = (j,q)$ or $\{i,j\} \in E$ and $\block(i,j) = \{(i,p),(j,q)\}$.
This will considerably simplify  the sum
\begin{equation}
    \ham[\tilde{t}]{aprx}_{i,j} = \sum_{p,q} \psi_p \psi_q \ham[\tilde{t}]{aprx}_{(i,p),(j,q)}.
    \label{eq:t-aprox-sum}
\end{equation}

\paragraph{Case 1.} Diagonal element: $i = j$.
First note that if $p = 0$ or $p > \degree_i$, then 
the block containing $(i,p)$ is just the single entry on the diagonal.
In this case, $\ham[r]{aprx}_{(i,p),(i,p)} = 1$ and
$\ham[\tilde{t}]{aprx}_{(i,p),(i,p)} = t_{(i,p),(i,p)}$.

Thus, when $i = j$, the sum (\ref{eq:t-aprox-sum})  simplifies to 
\begin{equation}
\ham[\tilde{t}]{aprx}_{i, i}
=
\frac12 t_{(i, 0), (i, 0)}
+
\frac1{2d}
\sum_{j: \{i, j\} \in E}
{\on\left(
\ham[R]{aprx}_{i, j} T_{i, j}
\ham[R]{aprx}_{i, j}
\right)}
+
\frac1{2d}
\sum_{p > \degree_i}
t_{(i, p), (i, p)}.
\end{equation}
The function $t$ is defined in (\ref{eq:t-alpha-def}) so that
$t_{\beta}(0) = t_{(i,0),(i,0)}$ and $t_{\alpha}(0) = t_{(i,p),(i,p)}$
for $p > 0$. Thus,
\begin{equation}
\ham[\tilde{t}]{aprx}_{i, i}
=
\frac12 t_{\beta}(0)
+
\frac1{2d}
\sum_{j: \{i, j\} \in E}
{\on\left(
\ham[R]{aprx}_{i, j} T_{i, j}
\ham[R]{aprx}_{i, j}
\right)}
+
\frac1{2d}
\sum_{p > \degree_i}
t_{\alpha}(0).
\end{equation}

Recall from (\ref{eq:t-round-def-on})
that the diagonal coefficients of $\ham[T]{round}$ are 
\begin{equation}
\ham[t]{round}_{i,i} = c_T
=
\frac12 \left(t_{\beta}(0) + t_{\alpha}(0)\right)
=
\frac12 t_{\beta}(0)
+
\frac1{2d}
\sum_{p > 0} t_{\alpha}(0).
\end{equation}
Therefore the difference between $\ham[\tilde{t}]{aprx}_{i,j}$ and $\ham[t]{round}_{i, j}$ is
\begin{align}
\left|
\ham[\tilde{t}]{aprx}_{i, i}
-
\ham[t]{round}_{i, i}
\right|
&=
\left|
\frac1{2d}
\sum_{j: \{i, j\} \in E}
{\on\left(
\ham[R]{aprx}_{i, j} T_{i, j}
\ham[R]{aprx}_{i, j}\right)-
t_{\alpha}(0)}
\right|
\\
&\leq
\frac1{2d}
\sum_{j: \{i, j\} \in E}
\left|
{\on\left(
\ham[R]{aprx}_{i, j} T_{i, j}
\ham[R]{aprx}_{i, j}
\right)
-
t_{\alpha}(0)}
\right|
\\
&\leq
\frac1{2d}
\sum_{j: \{i, j\} \in E}
\alpha \omega_{i, j}
\epsilon_{i, j}^2
& \mbox{by (\ref{eq:RTR-diag-bounds})}
\\
&\leq
\frac{\alpha}{2d}
\sum_{j: \{i, j\} \in E}
f(\omega_{i, j})
\\
&\leq
\frac{\alpha}{2d}
\cdot
d
\cdot
f(\omega_{\min})
    & \omega_{\min} \geq 2
\\
&
=
\frac{\alpha}{2}
f(\omega_{\min})
\leq
\alpha
f(\omega_{\min})
.
\label{eq:t-block-t-round-diff-diag}
\end{align}
Note that since $\omega_{\min} \ge 2$ (by the assumptions of the lemma), the function $f(\omega) = \omega^2 \exp(-\omega)$ is maximized at $\omega_{\min}$.

\paragraph{Case 2.} Off-diagonal element corresponding to edge: $\{i, j\} \in E$.
In this case, there is exactly one $p$ and exactly one $q$ such that $(i, p)$ and $(j, q)$ are in the same block,
where $\block(i,j) = \{(i,p),(j,q)\}$.
Thus, the summation in ~\cref{eq:t-aprox-sum} has only one non-zero term:
\begin{align}
\ham[\tilde{t}]{aprx}_{i, j}
=
\frac1{2d}
{\off\left(
\ham[R]{aprx}_{i, j}
T_{i, j}
\ham[R]{aprx}_{i, j}
\right)}
.
\end{align}
Recall from (\ref{eq:t-round-def-off}) that 
$
\ham[t]{round}_{i, j}
= 
-
\frac{\alpha}{4d}
\sqrt{f(\omega_{i, j})}
$.

Therefore
\begin{align}
\left|
\ham[\tilde{t}]{aprx}_{i, j}
-
\ham[t]{round}_{i, j}
\right|
    &=
\frac1{2d}
\left|
{\off\left(
\ham[R]{aprx}_{i, j}
T_{i, j}
\ham[R]{aprx}_{i, j}
\right)}
-
\left(
-\frac{\alpha}{2}
\sqrt{f(\omega_{i, j})}
\right)
\right|
\\
&\leq 
\frac1{2d}
\cdot
\frac{\alpha}{2}
\sqrt{f(\omega_{i, j})}
\cdot
4 \epsilon_{i, j}^2
& \mbox{by (\ref{eq:RTR-off-diag-bounds})}
\\
& = \frac{\alpha}{d} \omega_{i,j} \exp(-3 \omega_{i,j}/2)\\
&\le \alpha \omega_{i,j} \exp(-\omega_{i,j})\\
&\le \alpha
f(\omega_{\min}).
\label{eq:t-block-t-round-diff-edge}
\end{align}
Again, we are using the fact that since $\omega_{\min} \ge 2$, the function $f(\omega) = \omega^2 \exp(-\omega)$ is maximized at $\omega_{\min}$.

\paragraph{Case 3.} Off-diagonal element corresponding to non-edge: $\{i, j\} \notin E$.
In this case, $(i, p)$ and $(j, q)$ are in different blocks for all $p, q$, and so the summation in~\cref{eq:t-aprox-sum} is empty.
That is, $\ham[\tilde{t}]{aprx}_{i, j} = 0$.
Recall that $\ham[t]{round}_{i, j}$ is also zero for $\{i, j\} \notin E$.

Finally, we can combine the bound for
$\left|
\tilde{t}_{i, j}
-
\ham[\tilde{t}]{aprx}_{i, j}
\right|$ from (\ref{eq:tilde-aprx}) and
the bound for $\left|
\ham[\tilde{t}]{aprx}_{i, j} - \ham[t]{round}_{i,j}
\right|$ from (\ref{eq:t-aprx-t-round-diff}):

\begin{align}
\norm{
\ham[T]{ES}
-
\ham[T]{round}
}
&
\leq
\norm{
\sum_{\substack{i, j \\ \sigma}}
\left(
\tilde{t}_{i, j}
-
\ham[t]{round}_{i, j}
\right)
\tilde{a}_{i, \sigma}^{\dagger}
\tilde{a}_{j, \sigma}
}
\\
&\leq
\sum_{\substack{i, j \\ \sigma}}
\left|
\tilde{t}_{i, j}
-
\ham[t]{round}_{i, j}
\right|
\norm{
\tilde{a}_{i, \sigma}^{\dagger}
\tilde{a}_{j, \sigma}
}
=
2
\sum_{i, j}
\left|
\tilde{t}_{i, j}
-
\ham[t]{round}_{i, j}
\right|
\\
&\leq
2
\sum_{i, j}
\left|
\tilde{t}_{i, j}
-
\ham[\tilde{t}]{aprx}_{i, j}
\right|
+
2
\sum_{i, j}
\left|
\ham[\tilde{t}]{aprx}_{i, j}
-
\ham[t]{round}_{i, j}
\right|
\\
&\leq
10 n^8 \beta \exp(-\alpha \Gamma^2/4) + 2
n^2
\alpha f(\omega_{\min})
.
\end{align}

We can apply the conditions of the lemma to simplify this expression.
The lower bound on $\alpha \Gamma^2$ implies that $\exp(-\alpha \Gamma^2/4)
\le (10n^6\beta)^{-1} \cdot \exp(-\omega_{\min})$.
Using the assumptions that $\alpha \ge 1$ and $\omega_{\min} \ge 1$:
$$10 n^8 \beta \exp(-\alpha \Gamma^2/4) \le n^2 \exp(-\omega_{\min})
\le n^2 \alpha (\omega_{\min})^2  \exp(-\omega_{\min}) = n^2
\alpha f(\omega_{\min}).$$
Recall that $f(\omega) = \omega^2 \exp(-\omega)$.
The final bound for the kinetic-energy difference is
\begin{equation}
\norm{
\ham[T]{ES}
-
\ham[T]{round}
} \le 3
n^2
\alpha f(\omega_{\min}).
    \label{eq:T-ES-T-round-diff}
\end{equation}

Next, we consider the terms for the potential energy. 
As with the kinetic-energy terms, we will approximate
$\tilde{U} = (R \otimes R) U (R \otimes R)$
by 
\begin{equation}
\ham[\tilde{U}]{aprx} = (\ham[R]{aprx} \otimes \ham[R]{aprx}) \ham[U]{block} (\ham[R]{aprx} \otimes \ham[R]{aprx}).
\end{equation}
The matrices are now $n^2(d+1)^2 \times n^2(d+1)^2$. We will use the fact that if $A$ and $B$ are $m \times m$ matrices,
then $\max(|AB|) \le m \max(|A|) \cdot \max(|B|)$.
Since $d+1 \le n$, we pick up a factor of at most $n^4$ every time this
principle is applied. 
We will bound $\max(|\tilde{U}-\ham[\tilde{U}]{aprx}|)$ in two stages.
First we bound
\begin{align}
\max\left(\left| \tilde{U} - (R \otimes R) \ham[U]{block} (R \otimes R) \right|\right)   
& = \max\left(\left|(R \otimes R) U (R \otimes R) - (R \otimes R) \ham[U]{block} (R \otimes R)\right|\right)\\
    & = \max\left(\left|(R \otimes R) (U-\ham[U]{block}) (R \otimes R) \right|\right)\\
    & \le n^8 (r_{\max})^4 \ham[u]{neg}_{\max}\\
    & \le n^8 \underbrace{2^4}_{\ref{eq:r-bound}} \underbrace{2 \beta^3/\Gamma}_{\ref{eq:u-neg-bound}} =
    32 n^8 \beta^3/\Gamma.
    \label{eq:u-tilde-bound1}
\end{align}
The next step is to bound
\begin{align}
& \max\left(\left|(R \otimes R) \ham[U]{block} (R \otimes R) - \ham[\tilde{U}]{aprx}\right|\right)\\
=~~ &
\max\left(\left|(R \otimes R) \ham[U]{block} (R \otimes R) -
(\ham[R]{aprx} \otimes \ham[R]{aprx}) \ham[U]{block} (\ham[R]{aprx} \otimes \ham[R]{aprx})\right|\right).
    \label{eq:u-tilde-bound2a}
\end{align}
If we substitute $R = \ham[R]{neg} + \ham[R]{aprx}$ in to the expression
$(R \otimes R) \ham[U]{block} (R \otimes R)$ and expand the product, we 
get the sum of $2^4$ terms:
\begin{equation}
(R \otimes R) \ham[U]{block} (R \otimes R) = \sum_{a,b,c,d \in \{\mbox{neg},\mbox{aprx}\}} (\ham[R]{a} \otimes \ham[R]{b}) \ham[U]{block} (\ham[R]{c} \otimes \ham[R]{d}).
\end{equation}
In bounding the difference from (\ref{eq:u-tilde-bound2a}),
we are left with the terms in which $a,b,c,d$ are not all equal to
``aprx'', so every remaining term will have at least one factor
of $\ham[R]{neg}$:
\begin{align}
|(R \otimes R) \ham[U]{block} (R \otimes R) - \ham[\tilde{U}]{aprx}|
&\le n^8 u_{\max} \sum_{x=0}^3 {4 \choose x} \left(\ham[r]{aprx}_{\max} \right)^x
\left(\ham[r]{neg}_{\max} \right)^{4-x}\\
&\le n^8 u_{\max} \cdot 15 \cdot \ham[r]{neg}_{\max} 
\left[ \max\left\{ \ham[r]{aprx}_{\max} , \ham[r]{neg}_{\max} \right\} \right]^3\\
&\le n^8 \underbrace{2 \beta^3}_{(\ref{eq:u-bound})} \cdot 15 \cdot
\underbrace{n^2 \exp\left[-(\alpha \Gamma^2 - \omega_{\min})/2\right]}_{(\ref{eq:r-neg-bound})} \cdot   \underbrace{\left(\frac 3 2\right)^3}_{(\ref{eq:r-block-bound}, \ref{eq:r-neg-bound-constant})} 
    \\
    &\le 102 n^{10} \beta^3 \exp\left[-(\alpha \Gamma^2 - \omega_{\min})/2\right]\\
    & \le 102 n^{10} \beta^3 \exp(-\alpha \Gamma^2/4).
\label{eq:u-tilde-bound2}
\end{align}
The last inequality uses the assumption from the lemma that
$\alpha \Gamma^2 \ge 2 \omega_{\min}$.
Putting the two bounds from (\ref{eq:u-tilde-bound1}) and (\ref{eq:u-tilde-bound2})
together, we get that:
\begin{align}
& \max \left( \left| \tilde{U} - \ham[\tilde{U}]{aprx} \right| \right) \\
& \le \max \left( \left| \tilde{U} - (R \otimes R) \ham[U]{block} (R \otimes R) \right| \right)
+ \max \left( \left| (R \otimes R) \ham[U]{block} (R \otimes R) - \ham[\tilde{U}]{aprx} \right| \right)\\
&\le 32 n^8 \beta^3/\Gamma + 102 n^{10} \beta^3 \exp(-\alpha \Gamma^2/4).
\label{eq:u-tilde-bound-tot}
\end{align}

Since the composite orbitals are superpositions of the primitive orbitals,
the potential-energy terms for the composite orbitals can be expressed as linear
combinations of the potential-energy terms for the primitive orbitals. 
Therefore
\begin{equation}
    u_{i,j,k,l} = \sum_{p,q,r,s, \in [d+1]} \psi_p \psi_q \psi_r \psi_s
    u_{[(i,p),(j,q)][(l,r),(l,s)]},
    \label{eq:comp-obital-pot}
\end{equation}
where the amplitudes $\psi$ are defined to be $\psi_0 = 1/\sqrt{2}$ and
$\psi_{p>0} = 1/\sqrt{2d}$.
The same definition for $\tilde{u}_{i,j,k,l}$ and
$\ham[\tilde{u}]{aprx}_{i,j,k,l}$ can be applied using the potential-energy terms
for the primitive orbitals defined in $\tilde{U}$ and $\ham[\tilde{U}]{aprx}$.
We can apply the bound from (\ref{eq:u-tilde-bound-tot}) to bound
the difference in the potential-energy terms for the composite orbitals:
\begin{align}
    |\tilde{u}_{i,j,k,l} - \ham[\tilde{u}]{aprx}_{i,j,k,l}|
   & = \left|  \sum_{p,q,r,s, \in [d+1]} \psi_p \psi_q \psi_r \psi_s
    \left( \tilde{u}_{[(i,p),(j,q)][(l,r),(l,s)]} -
    \ham[\tilde{u}]{aprx}_{[(i,p),(j,q)][(l,r),(l,s)]} \right) \right|\\
    & \le \frac 1 4 (d+1)^4 \max \left( \left| \tilde{U} - \ham[\tilde{U}]{aprx} \right| \right)\\
    & \le \frac 1 4 n^4 (32 n^8 \beta^3/\Gamma + 102 n^{10} \beta^3 \exp(-\alpha \Gamma^2/4))\\
    & \le 8 n^{12} \beta^3/\Gamma + 26 n^{14} \beta^3 \exp(-\alpha \Gamma^2/4).
    \end{align}
    We can now apply the assumptions of the lemma to simplify
    the above expression.
    The assumption that $\Gamma \ge 640 n^{18} \beta^3$
    implies that $8 n^{12} \beta^3/\Gamma \le 1/(80 n^6)$.
    The assumption that $\alpha \Gamma^2 \ge 12 \log \beta + 80 \log n + 4 \omega_{\min} + 24 \geq 12 \log \beta + 80 \log n + 40$
    implies that $26 n^{14} \beta^3 \exp(-\alpha \Gamma^2/4) \le 1/(80 n^6)$. Therefore
\begin{equation}
    8 n^{12} \beta^3/\Gamma + 26 n^{14} \beta^3 \exp(-\alpha \Gamma^2/4)
    \le \frac{1}{40 n^6}.
    \label{eq:u-approx-1}
\end{equation}
The next task is to bound $|\ham[\tilde{u}]{aprx}_{i,j,k,l} - \ham[u]{round}_{i,j,k,l}|$.
Since $$\ham[\tilde{U}]{aprx} = (\ham[R]{aprx} \otimes \ham[R]{aprx}) \ham[U]{block} (\ham[R]{aprx} \otimes \ham[R]{aprx})$$
is block diagonal, many of the terms in the sum (\ref{eq:comp-obital-pot})
will be zero. We consider three cases. In each case, we will show
that
\begin{equation}
    |\ham[\tilde{u}]{aprx}_{i,j,k,l} - \ham[\tilde{u}]{round}_{i,j,k,l}| \le 4 \sqrt{\alpha} \cdot \exp{(-\omega_{\min}/2)}.
\label{eq:u-block-u-round-diff}
\end{equation}

\paragraph{Case 1.} Onsite term $i = j = k = l$.
Note that the entry in row $[(i,p),(i,q)]$ and row
$[(i,r),(i,s)]$ is outside of a block unless $p = q = r = s$.
If $p = 0$ or $p > \degree_i$, then 
the block containing $[(i,p),(i,p)]$ is just the single entry on the diagonal.
In this case, $\ham[r]{aprx}_{(i,p),(i,p)} \otimes \ham[r]{aprx}_{(i,p),(i,p)}= 1$ and the diagonal element at $[(i,p),(i,p)]$ is the same for
$\ham[\tilde{U}]{aprx}$ and $U$.
If $p=0$, then primitive orbital $\phi_{i,0}$ is a Gaussian of width $\beta$
and the diagonal term of $U$ at $[(i,0),(i,0)]$
is as in~\cref{eq:coulomb-integral} defined as $\ham[u]{Coul}_{\beta}(0)$.
For $p>\degree_i$, then primitive orbital $\phi_{i,p}$ is a Gaussian of width $\alpha$
and the diagonal term of $U$ at $[(i,p),(i,p)]$
is as in~\cref{eq:coulomb-integral} defined as $\ham[u]{Coul}_{\alpha}(0)$.

Thus, when $i = j  = k = l$, the sum (\ref{eq:comp-obital-pot})  simplifies to

\begin{align}
\ham[\tilde{u}]{aprx}_{i, i, i, i}
  & =
\frac1{4d^2}
\sum_{0 < p \leq \degree_i}
{\left(
{\left(\ham[R]{block}_{i, j}\right)}^{\otimes 2}
U_{i, j}
{\left(\ham[R]{block}_{i, j}\right)}^{\otimes 2}
\right)}_{(i, p), (i, p), (i, p), (i, p)} \\
& +
\frac14 
\ham[u]{Coul}_{\beta}(0)
+ 
\frac1{4d^2}
\sum_{p > \degree_i}
\ham[u]{Coul}_{\alpha}(0).
\end{align}
Recall that $\ham[u]{round}_{i,i,i,i}$ is defined in (\ref{eq:u-round-coul}) to be
\begin{align}
\ham[c]{round}_U 
&= 
\frac1{4}
\ham[u]{Coul}_{\beta}(0)
+
\frac1{4d}
\ham[u]{Coul}_{\alpha}(0)
= \frac1{4}
\ham[u]{Coul}_{\beta}(0)
+
\frac1{4d^2} \sum_{p \in[d]}
\ham[u]{Coul}_{\alpha}(0)
\end{align}
Therefore,
\begin{align}
&
\left|
\ham[\tilde{u}]{aprx}_{i, i, i, i}
-
\ham[u]{round}_{i, i, i, i}
\right|
\\
&=
\frac{1}{4d^2}
\left|
\sum_{0 < p \leq \degree_i}
\left(
{\left(
{\left(\ham[R]{block}_{i, j}\right)}^{\otimes 2}
U_{i, j}
{\left(\ham[R]{block}_{i, j}\right)}^{\otimes 2}
\right)}_{(i, p), (i, p), (i, p), (i, p)}
\right)
-
\ham[u]{Coul}_{\alpha}(0)
\right|
\\
&=
\frac{1}{4d^2}
\left|
\sum_{0 < p \leq \degree_i}
{\left(
{\left(\ham[R]{block}_{i, j}\right)}^{\otimes 2}
U_{i, j}
{\left(\ham[R]{block}_{i, j}\right)}^{\otimes 2}
-
U_{i, j}
\right)}_{(i, p), (i, p), (i, p), (i, p)}
\right|
\\
&\leq
\frac{1}{4d^2} \cdot d \cdot \max \left( \left|
{\left(\ham[R]{block}_{i, j}\right)}^{\otimes 2}
U_{i, j}
{\left(\ham[R]{block}_{i, j}\right)}^{\otimes 2}
-
U_{i, j}
\right)
\right|
\\
&\leq 
\frac14
\underbrace{16\sqrt{\alpha} \epsilon_{\max}}_{\labelcref{eq:RRURR-bound}}
=
4 \sqrt{\alpha} \cdot \exp(-\omega_{i,j}/2)
\leq
4 \sqrt{\alpha} \cdot \exp(-\omega_{\min}/2)
.
\end{align}
The last inequality uses the assumption of the lemma that $\omega_{\min} \ge 2$.

\paragraph{Case 2.} All indices within block corresponding to edge $\{i, j\} \in E$.
Let $\block(i,j) = \{(i,p),(j,q)\}$.

Consider, for example, the term $\ham[\tilde{u}]{aprx}_{i, j, j, i}$.
The sum in (\ref{eq:comp-obital-pot}) has only one non-zero term
corresponding to row $[(i,p),(j,q)]$ and column
$[(j,q),(i,p)]$. So
\begin{equation}
\ham[\tilde{u}]{aprx}_{i, j, j, i}
=
\frac1{4d^2}
{\left(
{\left(
\ham[R]{block}_{i, j}
\right)}^{\otimes 2}
U_{i, j}
{\left(
\ham[R]{block}_{i, j}
\right)}^{\otimes 2}
\right)}_{[(i, p), (j, q)], [(j, q), (i, p)]}
\end{equation}

Recall that
\begin{align}
\ham[u]{round}_{i, j, j, i}
&= 
\frac1{4d^2} \ham[u]{Coul}_{\alpha}(\gamma_{i, j})
=
\frac1{4d^2} 
u_{[(i, p), (j, q)], [(j, q), (i, p)]}
\end{align}
The first equality comes from the definition of $\ham[u]{round}$
in (\ref{eq:u-round-coul-diff}) and the second comes from the definition of
$\ham[u]{Coul}_{\alpha}(\gamma_{i, j})$ in (\ref{eq:coulomb-integral}).
The entry in row $[(i,p),(j,q)]$ and column
$[(j,q),(i,p)]$ is inside the block corresponding to edge $\{i,j\}$.
Therefore,
\begin{align}
\left|
\ham[\tilde{u}]{aprx}_{i, j, j, i}
-
\ham[u]{round}_{i, j, j, i}
\right|
&=
\frac1{4d^2}
\max \left( \left|
{\left(
\ham[R]{block}_{i, j}
\right)}^{\otimes 2}
U_{i, j}
{\left(
\ham[R]{block}_{i, j}
\right)}^{\otimes 2}
-
U_{i, j} \right|
\right)\\
& \leq
\frac14
\underbrace{16\sqrt{\alpha} \epsilon_{i,j}}_{\labelcref{eq:RRURR-bound}}
=
4 \sqrt{\alpha} \cdot \exp(-\omega_{i,j}/2)
\leq
4 \sqrt{\alpha} \cdot \exp(-\omega_{\min}/2)
.
\end{align}
The same bound holds for 
$\left|
\ham[\tilde{u}]{aprx}_{i, i, j, i}
-
\ham[u]{round}_{i, i, j, i}
\right|$
,
$\left|
\ham[\tilde{u}]{aprx}_{j, i, j, i}
-
\ham[u]{round}_{j, i, j, i}
\right|$,
et cetera for $\{i, j\} \in E$.

\paragraph{Case 3.} At least one pair of indices corresponding to non-edge $\{i, j\} \notin E$.
In this case, both $\ham[\tilde{u}]{aprx}_{i, j, k, l}$ and $\ham[u]{round}_{i, j, k, l}$ are zero.

All together,
\begin{align}
\norm{\ham[U]{ES} - \ham[U]{round}}
&\leq
\norm{
\frac12
\sum_{\substack{i, j, k, l \\ \sigma, \tau}}
\left(
\tilde{u}_{i, j, k, l}
-
\ham[u]{round}_{i, j, k, l}
\right)
\tilde{a}_{i, \sigma}^{\dagger}
\tilde{a}_{j, \sigma}^{\dagger}
\tilde{a}_{k, \sigma}
\tilde{a}_{l, \sigma}
}
\\
&\leq
\frac12
\sum_{\substack{i, j, k, l \\ \sigma, \tau}}
\left|
\tilde{u}_{i, j, k, l}
-
\ham[u]{round}_{i, j, k, l}
\right|
\norm{
\tilde{a}_{i, \sigma}^{\dagger}
\tilde{a}_{j, \sigma}^{\dagger}
\tilde{a}_{k, \sigma}
\tilde{a}_{l, \sigma}
}
\\
&=
\frac12 \cdot \underbrace{4}_{\sigma, \tau}
\sum_{i, j, k, l}
\left|
\tilde{u}_{i, j, k, l}
-
\ham[u]{round}_{i, j, k, l}
\right|
\\
&\leq
2
\sum_{i, j, k, l}
\left|
\tilde{u}_{i, j, k, l}
-
\ham[\tilde{u}]{aprx}_{i, j, k, l}
\right|
+
2
\sum_{i, j, k, l}
\left|
\ham[\tilde{u}]{aprx}_{i, j, k, l}
-
\ham[u]{round}_{i, j, k, l}
\right|
\\
&\leq
2n^4
\underbrace{ \left( \frac{1}{40n^6} \right) }_{(\ref{eq:u-approx-1})}
+
2n^4
\cdot
\underbrace{4 \sqrt{\alpha} \cdot \exp(-\omega_{\min}/2)}_{\labelcref{eq:u-block-u-round-diff}}\\
& = \frac{1}{20n^2}  + 8 n^4 \sqrt{\alpha} \cdot \exp(-\omega_{\min}/2)
\label{eq:U-ES-U-round-diff}
\end{align}
\cref{eq:T-ES-T-round-diff,eq:U-ES-U-round-diff} imply the lemma:
\begin{align}
\norm{\ham{ES} - \ham{round}}
&\leq
\norm{\ham[T]{ES} - \ham[T]{round}}
+
\norm{\ham[U]{ES} - \ham[U]{round}}
\\
&\leq
3 n^2 \alpha f(\omega_{\min})
+
\frac{1}{20n^2}  + 8 n^4 \sqrt{\alpha} \cdot \exp(-\omega_{\min}/2)
.
\end{align}
\end{proof}

\subsection{Proof of~\cref{lem:r-bound}}

\begin{proof}[Proof of~\cref{lem:r-bound}]
Define $\epsilon_{\max}$ to be the largest off-diagonal element of
    $S$ and $\epsilon_{\mathrm{neg}}$ to be the largest entry of $\ham[S]{neg}$,
which is the largest entry of $S$ outside of an edge block:
\begin{align}
\epsilon_{\max} &= 
\max_{(i,p) \neq (j,q)}
s_{(i,p), (j,q)}
=
s_{\alpha} (\gamma_{\min}) = \exp(-\omega_{\min} / 2),
\\
\epsilon_{\mathrm{neg}}
&=
\max\left(
\ham[S]{neg} \right)
\leq 
s_{\alpha} (\Gamma)
=
\exp(-\alpha \Gamma^2/2)
.
\end{align}
Let $\ham[S]{block} = I + \ham[S]{OD}$; $\ham[S]{OD}$ has at most one entry per row or column, and that entry is between $0$ and $\epsilon_{\max}$.
Using the Taylor expansion
\begin{align}
{
M
}^{-1/2}
=
\sum_{k=0}^{\infty}
{(-2)}^{-k}
\frac{(2k-1)!!}{k!}
{\left(
M - I
\right)}^k
\end{align}
of a matrix $M$ around the identity $I$, we have
\begin{align}
\ham[R]{neg}
&=
R
-
\ham[R]{aprx}
\\
&=
{\left(
S
\right)}^{-1/2}
-
{\left(
\ham[S]{block}
\right)}^{-1/2}
\\
&=
{\left(
I + \ham[S]{OD} + \ham[S]{neg}
\right)}^{-1/2}
-
{\left(
I + \ham[S]{OD}
\right)}^{-1/2}
\\
&=
\sum_{k=0}^{\infty}
{(-2)}^{-k}
\frac{(2k-1)!!}{k!}
\left[
{\left(
\ham[S]{OD} + \ham[S]{neg}
\right)}^k
-
{\left(
\ham[S]{OD}
\right)}^k
\right]
.
\label{eq:R-neg-expansion}
\end{align}
Entrywise,
\begin{align}
&
{\left[
{\left(
\ham[S]{OD} + \ham[S]{neg}
\right)}^k
\right]}_{(i_0, l_0), (i_k, l_k)}
\\
&=
\sum_{\mathclap{(i_1, l_1), \ldots, (i_{k-1}, l_{k-1})}}
{\left(
\ham[S]{OD} + \ham[S]{neg}
\right)}_{(i_0, l_0), (i_1, l_1)}
\cdots
{\left(
\ham[S]{OD} + \ham[S]{neg}
\right)}_{(i_{k-1}, l_{k-1}), (i_k, l_k)}
\\
&=
\sum_{\mathclap{
0 < \norm{
\mathbf x_{i_{k'}, l_{k'}} -  \mathbf x_{i_{k'+1}, l_{k'+1}}
}
}}
s_{(i_0, l_0), (i_1, l_1)}
\cdots
s_{(i_{k-1}, l_{k-1}), (i_k, l_k)}
,
\label{eq:S-power}
\end{align}
where the summation excludes the diagonal entries.
(Recall that $I$, $\ham[S]{OD}$, and $\ham[S]{neg}$ have disjoint support.)
Similarly,
\begin{align}
{\left[
{\left(
\ham[S]{OD}
\right)}^k
\right]}_{(i_0, l_0), (i_k, l_k)}
&=
\sum_{{
0 < \norm{
\mathbf x_{i_{k'}, l_{k'}} -  \mathbf x_{i_{k'+1}, l_{k'+1}}
}
< \Gamma
}}
s_{(i_0, l_0), (i_1, l_1)}
\cdots
s_{(i_{k-1}, l_{k-1}), (i_k, l_k)}
,
\label{eq:S-OD-power}
\end{align}
where the summation excludes both the diagonal and anything outside of the blocks.
The difference between~\cref{eq:S-power} and~\cref{eq:S-OD-power} is the summation in~\cref{eq:S-power} restricted to when at least one of the neighboring pairs is at least $\Gamma$ separated.
Each term with exactly $x$ pairs separated by at least $\Gamma$ contributes at most $\epsilon_{\max}^{k-x} \epsilon_{\mathrm{neg}}^x$.
There are $\binom{k}{x}$ places in the sequence that these pairs can occur.
For each factor contributing more than $\epsilon_{\mathrm{neg}}$ there is at most one index value $(i_{k'}, l_{k'})$, and for each factor contributing at most $\epsilon_{\mathrm{neg}}$ there are at most $n(d+1) - 1 \leq 2n^2$ indices.
Therefore
\begin{align}
&
{\left[
{\left(
\ham[S]{OD} + \ham[S]{neg}
\right)}^k
-
{\left(
\ham[S]{OD}
\right)}^k
\right]}_{(i_0, l_0), (i_k, l_k)}
\\
&
\leq
\sum_{x=1}^k
\binom{k}{x}
{\left(
2n^2 \epsilon_{\mathrm{neg}}
\right)}^x
\epsilon_{\max}^{k-x}
\\
&
=
\sum_{x=0}^{k-1}
\binom{k}{x}
{\left(
2n^2 \epsilon_{\mathrm{neg}}
\right)}^{k-x}
\epsilon_{\max}^{x}
=
{\left(
2n^2 \epsilon_{\mathrm{neg}}
\right)}^{k}
\sum_{x=0}^{k-1}
\frac{k}{k-x}
\binom{k-1}{x}
{\left(
\frac{\epsilon_{\max}}{
2n^2 \epsilon_{\mathrm{neg}}
}
\right)}^{x}
\\
&\leq
{\left(
2n^2 \epsilon_{\mathrm{neg}}
\right)}^{k}
\sum_{x=0}^{k-1}
k
\binom{k-1}{x}
{\left(
\frac{\epsilon_{\max}}{
2n^2 \epsilon_{\mathrm{neg}}
}
\right)}^{x}
=
k
{\left(
2n^2 \epsilon_{\mathrm{neg}}
\right)}^{k}
{\left(
1+
\frac{\epsilon_{\max}}{
2n^2 \epsilon_{\mathrm{neg}}
}
\right)}^{k-1}
\\
&=
k
{\left(
2n^2 \epsilon_{\mathrm{neg}}
\right)}
{\left(
2n^2 \epsilon_{\mathrm{neg}}
+
\epsilon_{\max}
\right)}^{k-1}
\\
&\leq
k
{\left(
2n^2 \epsilon_{\mathrm{neg}}
\right)}
{\left(
2 \epsilon_{\max}
\right)}^{k-1}
=
kn^2 \frac{\epsilon_{\mathrm{neg}}}{\epsilon_{\max}}
{\left(
2\epsilon_{\max}
\right)}^k
,
\end{align}
where we used the fact that
\begin{equation}
2n^2 \epsilon_{\mathrm{neg}}
=
2
n^2 e^{-\alpha \Gamma^2 / 2}
\leq
e
\cdot
n^2 e^{-\alpha \Gamma^2 / 2}
\leq 
e
\cdot
n^2 e^{-(4 \log n + \omega_{\min} + 2)/ 2}
=
e^{-\omega_{\min} / 2}
=
\epsilon_{\max}
\end{equation}
by assumption.
Returning to the expression in~\cref{eq:R-neg-expansion}, the norm of each entry of $\ham[R]{neg}$ then is
\begin{align}
\left|
\ham[r]{neg}_{(i_0, l_0), (i_k, l_k)}
\right|
&=
\left|
\sum_{k=0}^{\infty}
{(-2)}^{-k}
\frac{(2k-1)!!}{k!}
{\left[
{\left(
\ham[S]{OD} + \ham[S]{neg}
\right)}^k
-
{\left(
\ham[S]{OD}
\right)}^k
\right]}_{(i_0, l_0), (i_k, l_k)}
\right|
\\
&\leq
\sum_{k=0}^{\infty}
{(2)}^{-k}
\frac{(2k-1)!!}{k!}
\left|
{\left[
{\left(
\ham[S]{OD} + \ham[S]{neg}
\right)}^k
-
{\left(
\ham[S]{OD}
\right)}^k
\right]}_{(i_0, l_0), (i_k, l_k)}
\right|
\\
& \leq
n^2
\frac{\epsilon_{\mathrm{neg}}}{\epsilon_{\max}}
\sum_{k=1}^{\infty}
{(2)}^{-k}
\frac{(2k-1)!!}{k!}
k
{\left(
2\epsilon_{\max}
\right)}^{k}
\\
&=
n^2
\frac{\epsilon_{\mathrm{neg}}}{\epsilon_{\max}}
\sum_{k=1}^{\infty}
\frac{(2k-1)!!}{(k-1)!}
\epsilon_{\max}^k
\\
&=
n^2
\frac{\epsilon_{\mathrm{neg}}}{\epsilon_{\max}}
\frac{
\epsilon_{\max}
}{
{(1-2\epsilon_{\max})}^{3/2}
}
\\
&\leq 
n^2 
\frac{\epsilon_{\mathrm{neg}}}{\epsilon_{\max}}
& \text{for $\omega_{\min} \geq 4$}\\
&\leq n^2 \exp\left[-(\alpha \Gamma^2 - \omega_{\min})/2\right]
.
\end{align}
\end{proof}

\subsection{Proof of~\cref{lem:r-approx}}

\begin{proof}[Proof of~\cref{lem:r-approx}]
In this proof, we'll use the following form of Taylor's Theorem.
\begin{theorem}[Taylor's Theorem with remainder in Lagrange form~\cite{apostol2007calculus}]
\label{thm:taylor}
Let $f$ be a $(n+1)$-times differentiable function in the region $[0, 1]$.
Then for every $x \in [0, 1]$ there is some $c \in [0, x]$ such that
\begin{equation}
f(x) = 
\sum_{k=0}^n
\frac{f^{(k)}(0)}{k!} x^k 
+
\frac{f^{(n+1)}(c)}{(n +1)!}
x^{n+1}
.
\end{equation}
\end{theorem}
Note that $\omega_{\min} \geq 2$ implies that $\epsilon_{i, j} = e^{-\omega_{i, j} / 2} \leq e^{-\omega_{\min} /2} \leq 1 /e$.

We'll start with the bounds on the entries of $\ham[R]{aprx}$.
Given the derivatives
\begin{align}
\frac{d}{d\epsilon}
\left(
\frac{1}{\sqrt{1 + \epsilon}} 
\pm
\frac{1}{\sqrt{1 - \epsilon}} 
\right)
&=
\frac12
\left(
-
{(1  + \epsilon)}^{-3/2}
\pm
{(1  - \epsilon)}^{-3/2}
\right)
,
\\
\frac{d^2}{d\epsilon^2}
\left(
\frac{1}{\sqrt{1 + \epsilon}} 
\pm
\frac{1}{\sqrt{1 - \epsilon}} 
\right)
&=
\frac34
\left(
{(1  + \epsilon)}^{-5/2}
\pm
{(1  - \epsilon)}^{-5/2}
\right)
,
\\
\frac{d^3}{d\epsilon^3}
\left(
\frac{1}{\sqrt{1 + \epsilon}} 
\pm
\frac{1}{\sqrt{1 - \epsilon}} 
\right)
&=
\frac{15}{8}
\left(
-
{(1  + \epsilon)}^{-7/2}
\pm
{(1  - \epsilon)}^{-7/2}
\right)
,
\end{align}
\cref{thm:taylor} implies that
\begin{align}
\frac{1}{\sqrt{1 + \epsilon}}
+
\frac{1}{\sqrt{1 - \epsilon}}
&=
2 + 0 + 
\frac12 
\frac34
\left(
{(1  + \epsilon')}^{-5/2}
+
{(1  - \epsilon')}^{-5/2}
\right)
\epsilon^2
,
\\
\frac{1}{\sqrt{1 + \epsilon}}
-
\frac{1}{\sqrt{1 - \epsilon}}
&=
0
-\epsilon
+
0
-
\frac{1}{3!}
\frac{15}8
\left(
{(1  + \epsilon')}^{-7/2}
+
{(1  - \epsilon')}^{-7/2}
\right)
\epsilon^3
\end{align}
for some $\epsilon' \in [0, \epsilon]$.
For $0 \leq \epsilon \leq 1 / e$, we have
\begin{align}
2 
&\leq 
\frac{1}{\sqrt{1 + \epsilon}}
+
\frac{1}{\sqrt{1 - \epsilon}}
\leq
2 + 2 \epsilon^2
,
\\
-\epsilon - 2 \epsilon^3
&\leq 
\frac{1}{\sqrt{1 + \epsilon}}
-
\frac{1}{\sqrt{1 - \epsilon}}
\leq
-\epsilon
.
\end{align}
Dividing by 2 and substituting $\epsilon = \epsilon_{i, j}$ implies~\cref{eq:r-block-diag-bounds,eq:r-block-off-diag-bounds}.

Now, let's turn to the entries of $\ham[R]{aprx}_{i, j} T_{i, j} \ham[R]{aprx}_{i, j}$.
Let $\block(i,j) = \{(i,p),(j,q)\}$.
To make the notation more concise within this proof, we
will refer to the diagonal elements of $\ham[R]{aprx}_{i, j}$
as $r_{ON} = \ham[r]{aprx}_{(i,p),(i,p)} = \ham[r]{aprx}_{(j,q),(j,q)}$
and the off-diagonal elements as
$r_{OFF} = \ham[r]{aprx}_{(i,p),(j,q)} = \ham[r]{aprx}_{(j,q),(i,p)}$.
Note that
\begin{align}
{\left(r_{ON}\right)}^2
+
{\left(r_{OFF}\right)}^2
&=
\frac14
\left(
{\left(
\frac1{\sqrt{1 + \epsilon_{i, j}}}
+
\frac1{\sqrt{1 - \epsilon_{i, j}}}
\right)}^2
+
{\left(
\frac1{\sqrt{1 + \epsilon_{i, j}}}
-
\frac1{\sqrt{1 - \epsilon_{i, j}}}
\right)}^2
\right)
\\
&=
\frac12
\left(
\frac1{1 + \epsilon_{i, j}}
+
\frac1{1 - \epsilon_{i, j}}
\right)
=
\frac{1}{1 - \epsilon_{i, j}^2}
\end{align}
and
\begin{align}
r_{ON} \cdot
r_{OFF}
&=
\frac14
\left(
\frac1{\sqrt{1 + \epsilon_{i, j}}}
+
\frac1{\sqrt{1 - \epsilon_{i, j}}}
\right)
\left(
\frac1{\sqrt{1 + \epsilon_{i, j}}}
-
\frac1{\sqrt{1 - \epsilon_{i, j}}}
\right)
\\
&=
\frac14
\left(
\frac1{1 + \epsilon_{i, j}}
-
\frac1{1 - \epsilon_{i, j}}
\right)
= -\frac{\epsilon_{i, j}}{2(1 - \epsilon_{i, j}^2)}
.
\end{align}
The diagonal entries of $T_{i,j}$ are
$t_{(i,p),(i,p)} = t_{(j,q),(j,q)} = t_{\alpha}(0)$.
and the off-diagonal entries are
$t_{(i,p),(j,q)} = t_{(j,q),(i,p)} = t_{\alpha}(\gamma_{i, j})$.
Then the diagonal of $\ham[R]{aprx}_{i, j} T_{i, j} \ham[R]{aprx}_{i, j}$ entry is
\begin{align}
&
{\on \left(
\ham[R]{aprx}_{i,j}
T_{i, j}
\ham[R]{aprx}_{i,j}
\right)}
\\
&=
\begin{pmatrix}
r_{ON} &
r_{OFF}
\end{pmatrix}
\begin{pmatrix}
t_{\alpha}(0) &
t_{\alpha}(\gamma_{i, j}) \\
t_{\alpha}(\gamma_{i, j}) &
t_{\alpha}(0) 
\end{pmatrix}
\begin{pmatrix}
r_{ON} \\
r_{OFF}
\end{pmatrix}
\\
&=
t_{\alpha}(0)
\left(
{\left(
r_{ON}
\right)}^2
+
{\left(
r_{OFF}
\right)}^2
\right)
+
2
t_{\alpha}(\gamma_{i, j})
r_{ON}
r_{OFF}
\\
&=
\frac{t_{\alpha}(0)}{1 - \epsilon_{i, j}^2}
-
\frac{t_{\alpha}(\gamma_{i, j}) \epsilon_{i, j}}{1 - \epsilon_{i, j}^2}
\\
&=
\left(
t_{\alpha}(0)
-
s_{\alpha}(\gamma_{i, j})
t_{\alpha}(\gamma_{i, j})
\right)
\frac{1}{1 - \epsilon_{i, j}^2}
\\
&=
\left(
\frac32 \alpha
-
\frac12 \alpha (3 - \omega_{i, j}) \exp(-\omega_{i, j})
\right)
\frac{1}{1 - \epsilon_{i, j}^2}
\\
&=
\left(
\frac32 \alpha
(1 - \epsilon_{i, j}^2)
+
\frac12 \alpha \omega_{i, j} \epsilon_{i, j}^2
\right)
\frac{1}{1 - \epsilon_{i, j}^2}
\\
&=
t_{\alpha}(0)
+ 
\frac{\alpha \omega_{i, j} \epsilon_{i, j}^2}{2 (1 - \epsilon_{i, j}^2)}
\label{eq:RTR-diag}
\end{align}
and the off-diagonal entry of $\ham[R]{aprx}_{i, j} T_{i, j} \ham[R]{aprx}_{i, j}$
is
\begin{align}
&
{\off \left(
\ham[R]{aprx}_{i, j}
T_{i, j}
\ham[R]{aprx}_{i, j}
\right)}
\\
&=
\begin{pmatrix}
r_{ON} &
r_{OFF}
\end{pmatrix}
\begin{pmatrix}
t_{\alpha}(0) & 
t_{\alpha}(\gamma_{i, j}) \\
    t_{\alpha}(\gamma_{i, j}) &
t_{\alpha}(0)
\end{pmatrix}
\begin{pmatrix}
r_{OFF} \\
r_{ON}
\end{pmatrix}
\\
&=
2 
t_{\alpha}(0) \cdot
r_{ON} \cdot
r_{OFF}
+
t_{\alpha}(\gamma_{i, j}) 
\left(
{\left(r_{ON}\right)}^2
+
{\left(r_{OFF}\right)}^2
\right)
\\
&=
-\frac{t_{\alpha}(0) \epsilon_{i, j}}{1 - \epsilon_{i, j}^2}
+
\frac{
t_{\alpha}(\gamma_{i, j}) 
}{
1 - \epsilon_{i, j}^2
}
\\
&=
\left(
t_{\alpha}(\gamma_{i, j}) 
- 
s_{\alpha}(\gamma_{i, j}) t_{\alpha}(0)
\right)
\frac1{1 - \epsilon_{i, j}^2}
\\
&=
\left(
\frac12 \alpha (3 - \omega_{i, j})
\exp(-\omega_{i, j} / 2)
- 
\frac32 \alpha \exp(-\omega_{i, j} / 2)
\right)
\frac1{1 - \epsilon_{i, j}^2}
\\
&=
-\frac12 \alpha \omega_{i, j} \exp(-\omega_{i, j} / 2)
\frac1{1 - \epsilon_{i, j}^2}
\\
&=
-\frac12 \alpha \sqrt{f(\omega_{i, j})}
\frac1{1 - \epsilon_{i, j}^2}
.
\label{eq:RTR-off-diag}
\end{align}
Let's look at this factor ${(1 - \epsilon^2)}^{-1}$.
It's always at least $1$, and its first two derivatives are
\begin{align}
\frac{d}{d \epsilon}
\left(\frac1{1 - \epsilon^2}\right)
&=
\frac{2 \epsilon }{{\left(1 - \epsilon^2\right)}^2}
,
\\
\frac{d^2}{d \epsilon^2}
\left(\frac1{1 - \epsilon^2}\right)
&=
\frac{2 (1 + 3 \epsilon^2)}{{\left(1 - \epsilon^2\right)}^3}
.
\end{align}
By~\cref{thm:taylor},
\begin{align}
\frac{1}{1 - \epsilon^2}
=
1 + 
\frac{(1 + 3 \epsilon'^2)}{{\left(1 - \epsilon'^2\right)}^3}
\epsilon^2
\end{align}
for some $ 0 \leq \epsilon' \leq \epsilon$.
For $\epsilon \leq 1 / e$, we have
\begin{align}
1 & \leq 
\frac1{1 - \epsilon^2}
\leq
1 + 4 \epsilon^2
\leq
2
.
\label{eq:factor-bounds}
\end{align}
Combining~\cref{eq:RTR-diag,eq:RTR-off-diag,eq:factor-bounds} implies~\cref{eq:RTR-diag-bounds,eq:RTR-off-diag-bounds}.

Finally, we turn to the bound for
$$
\max \left|
{\left(
\ham[R]{aprx}_{i, j}
\right)}^{\otimes 2}
U_{i, j}
{\left(
\ham[R]{aprx}_{i, j}
\right)}^{\otimes 2}
-
U_{i, j} \right|
$$
Each entry of $\left(
\ham[R]{aprx}_{i, j}
\right)^{\otimes 2}$
is a product of two terms from $\{ r_{ON}, r_{OFF}\}$, and only the diagonal
terms are $(r_{ON})^2$.
For notational ease, we will index the  four rows and columns 
of $U_{i,j}$ by $\{0,1,2,3\}$.
For $a, b \in \{0,1,2,3\}$,
we will denote the entry in row $a$ and column $b$ by $U_{i,j}[a,b]$.
Now consider a particular entry in row $a$ and column $b$ of ${\left(
\ham[R]{aprx}_{i, j}
\right)}^{\otimes 2}
U_{i,j}
{\left(
\ham[R]{aprx}_{i, j}
\right)}^{\otimes 2}$.
This entry is the sum of $16$ terms, each of which is a product of one entry from $U_{i,j}$
and four factors from $\{ r_{ON}, r_{OFF}\}$.
The only term that has four factors of $(r_{ON})$ is $(r_{ON})^4 U_{i,j}[a,b]$
because the two factors of $(r_{ON})^2$
must come from diagonal entries of
${\left(
\ham[R]{aprx}_{i, j}
\right)}^{\otimes 2}$.
The other $15$ terms all have at least one factor of $r_{OFF}$.
Also since
$|r_{OFF}| < |r_{ON}|$, each of these other terms is at most 
$|r_{OFF}| |r_{ON}|^3 \cdot \max\left(U_{i,j} \right)$.
Therefore we have:
\begin{align}
  &  \left|
{\left(
\ham[R]{aprx}_{i, j}
\right)}^{\otimes 2}
U_{i, j}
{\left(
\ham[R]{aprx}_{i, j}
\right)}^{\otimes 2}[a,b]
-
U_{i, j}[a,b] \right|\\
\le &
\left| (r_{ON})^4 U_{i, j}[a,b] + 15 \cdot | r_{OFF}| |r_{ON}|^3 \max\left(U_{i,j} \right)
- U_{i, j}[a,b] \right|
\end{align}
The maximum entry in $U_{i,j}$ is $\ham[u]{Coul}_{\alpha}(0)$. 
Therefore 
$$
\max \left|
{\left(
\ham[R]{aprx}_{i, j}
\right)}^{\otimes 2}
U_{i, j}
{\left(
\ham[R]{aprx}_{i, j}
\right)}^{\otimes 2}
-
U_{i, j} \right| \le \ham[u]{Coul}_{\alpha}(0)
\left[(r_{ON})^4 - 1 + 15 \cdot |r_{OFF}| \cdot |r_{ON}|^3\right]
$$
Using the bounds from ~\cref{eq:r-block-diag-bounds,eq:r-block-off-diag-bounds},
we know that $|r_{OFF}| \le \epsilon_{i,j}(1/2 + \epsilon_{i,j}^2)$
and $|r_{ON}| \le 1 + \epsilon_{i,j}^2$.
Also $\epsilon_{i,j} = \exp(\omega_{i,j}/2)$ and since by assumption
$\omega_{i,j} \ge 4$, $\epsilon \le 1/4$.

\begin{align}
  &  \max \left|
{\left(
\ham[R]{aprx}_{i, j}
\right)}^{\otimes 2}
U_{i, j}
{\left(
\ham[R]{aprx}_{i, j}
\right)}^{\otimes 2}
-
U_{i, j} \right|\\
\le & \ham[u]{Coul}_{\alpha}(0)
\left[(r_{ON})^4 - 1 + 15 |r_{OFF}| \cdot |r_{ON}|^3\right]\\
\le & \ham[u]{Coul}_{\alpha}(0)
\left[(1 + \epsilon_{i,j}^2)^4 - 1 + 15 \epsilon_{i,j}(1/2 + \epsilon_{i,j}^2) \cdot (1 + \epsilon_{i,j}^2)^3\right]\\
\le & \ham[u]{Coul}_{\alpha}(0)
\left[2\epsilon_{i,j} +  15 \left(\frac{9}{16}\right) \cdot \left(\frac{17}{16}\right)^3 \epsilon_{i,j}\right]\\
\le & \ham[u]{Coul}_{\alpha}(0) \epsilon_{i,j} \cdot 12 = \frac{2}{\sqrt{\pi}} 13 \sqrt{\alpha}
\epsilon_{i,j} \le 16  \sqrt{\alpha}
\epsilon_{i,j}
\end{align}

\end{proof}

\subsection{Proof of~\cref{lem:integral-bounds}}

\begin{proof}[Proof of~\cref{lem:integral-bounds}]
We'll start with the kinetic coefficient bounds.
Each $t_{(i, p), (j, q)}$ coefficient has one
of the following three forms (from \cref{eq:gen-t-integral}), depending
on whether orbitals $\phi_{i,p}$ and $\phi_{j,q}$ have exponent
$\alpha$ or $\beta$, and where $x$ is the distance
between the two orbitals:
\begin{align}
t_{\alpha, \beta}(x) 
&=
2^{3/2} 
\frac{{(\alpha \beta)}^{7/4}}{{(\alpha + \beta)}^{5/2}}
\left(3 - 2 \mu x^2\right)
\exp\left(-\mu x^2\right) ~~~~~~~ \mbox{where }\mu = \alpha \beta/(\alpha+\beta)\\
t_{\alpha}(x) 
&=
\frac{\alpha}{2}
\left(3 - \alpha x^2\right)
\exp\left(-\alpha x^2/2\right)\\
t_{\beta}(x) 
&=
\frac{\beta}{2}
\left(3 - \beta x^2\right)
\exp\left(-\beta x^2/2\right)
\end{align}
Consider the prefactor:
\begin{align}
2^{3/2} 
\frac{{(\alpha \beta)}^{7/4}}{{(\alpha + \beta)}^{5/2}}
= \frac{\sqrt{\alpha \beta}}{2} \left[ \frac{\sqrt{\alpha \beta}}{\left( \frac{\alpha+\beta}{2} \right)}\right]^{5/2} \le \frac{\sqrt{\alpha \beta}}{2}
\end{align} 
The  inequality follows from the fact that the geometric mean of two positive numbers is no more than their arithmetic mean.
Therefore, since $\beta \ge \alpha$, the maximum prefactor for $t_{\alpha}(x)$, $t_{\beta}(x)$,
or $t_{\alpha,\beta}(x)$ is
\begin{equation}
\label{eq:prefactor-bound}
\max\left\{ \frac{\alpha}{2}, \frac{\beta}{2}, \frac{\sqrt{\alpha \beta}}{2} \right\} = \frac{\beta}{2}
.
\end{equation}

The part of  the function $t$ that depends on $x$ is
$$\bar{t}_{\mu}(x) = (3 - 2 \mu x^2) \exp(\mu x^2),$$
where $\mu = \alpha \beta/(\alpha+\beta)$
or $\alpha/2$ or $\beta/2$.
Note that $\bar{t}_{\mu} (x)$ changes sign once, from positive to negative, at $2 \mu x^2 = 3$.
Its derivative,
\begin{align}
\bar{t}'_{\mu}(x) 
&=
2^{3/2} 
\mu x (4 \mu x^2 - 10)
\exp\left(-\mu x^2\right)
,
\end{align}
vanishes only at the origin and $2 \mu x^2 = 5$, where it goes from negative to positive.
Therefore, 
\begin{equation}
\max_{x \geq 0}
\left|\bar{t}_{\mu}(x)\right| 
=
\max\left\{
\bar{t}_{\mu}(0),
-\bar{t}_{\mu}(\sqrt{5/2\mu}) 
\right\}
=
\bar{t}_{\mu}(0) 
=
3.
\end{equation}
Putting this together with the bound on the prefactor from
(\ref{eq:prefactor-bound}), we get that 
$t_{\max} \leq \frac32 \beta$.

$\mu \ge \alpha/2$, and therefore
$2 \mu \Gamma^2 \geq \alpha \Gamma^2$.
Since, by assumption, $\alpha \Gamma^2 \ge 5$, we know that
$\bar{t}_{\mu}(x)$ is monotonic for $x \geq \Gamma$. Therefore
\begin{align}
\max_{x \geq \Gamma}
\left|
\bar{t}_{\mu}(x)
\right|
&=
\left|
\bar{t}_{\mu}(\Gamma)
\right|
\\
&=
\left|3 - 2 \mu \Gamma^2\right|
\exp\left(-\mu \Gamma^2\right)
\\
&\leq
2 \mu \Gamma^2
\exp\left(-\mu \Gamma^2\right)
\\
&\leq
2 \exp\left(-\mu \Gamma^2 / 2\right)
\leq
2
\exp\left(-\alpha \Gamma^2 / 4\right)
,
\end{align}
where in getting to the last line we used that $x e^{-x} \leq e^{-x/2}$.
Putting this together with the bound on the prefactor from
(\ref{eq:prefactor-bound}), we get that  $\ham[t]{neg}_{\max} \leq 
\beta
\exp\left(-\alpha \Gamma^2 / 4\right)
$.

Bounding the potential integrals will be easier because the integrand is strictly positive.
Each potential integral corresponds to four Gaussians with centers $\mathbf x_1$ through $\mathbf x_4$ and exponents $\zeta_1$ through $\zeta_4$:
\begin{align}
&
\int d \mathbf r d \mathbf s
\xi_{\zeta_1}(\mathbf r - \mathbf x_1)
\xi_{\zeta_2}(\mathbf s - \mathbf x_2)
    \frac{1}{\norm{\mathbf r - \mathbf s}}
\xi_{\zeta_3}(\mathbf s - \mathbf x_3)
\xi_{\zeta_4}(\mathbf r - \mathbf x_4)
\\
&=
\prod_{i=1}^4
{\left(
\frac{2\zeta_i}{\pi}
\right)}^{3/4}
\int d \mathbf r d \mathbf s
\exp\left[
-\zeta_1 \norm{\mathbf r - \mathbf x_1}^2
-\zeta_2 \norm{\mathbf s - \mathbf x_2}^2
-\zeta_3 \norm{\mathbf s - \mathbf x_3}^2
-\zeta_4 \norm{\mathbf r - \mathbf x_4}^2
\right]
\frac{1}{\norm{\mathbf r - \mathbf s}}
\\
&\leq
{\left(
\frac{2\beta}{\pi}
\right)}^3
\int d \mathbf r d \mathbf s
\exp\left[
-\alpha \left(
\norm{\mathbf r - \mathbf x_1}^2
+
\norm{\mathbf s - \mathbf x_2}^2
+
\norm{\mathbf s - \mathbf x_3}^2
+
\norm{\mathbf r - \mathbf x_4}^2
\right)
\right]
\frac{1}{\norm{\mathbf r - \mathbf s}}
\\
&=
{\left(
\frac{2\beta}{\pi}
\right)}^3
\int d \mathbf r d \mathbf s
\exp\left[
-\alpha \left(
2 \norm{\mathbf r - \frac{\mathbf x_1 + \mathbf x_4}{2}}^2
+
\frac12 \norm{\mathbf x_1 - \mathbf x_4}^2
+
2 \norm{\mathbf s - \frac{\mathbf x_2 + \mathbf x_3}{2}}^2
+
\frac12 \norm{\mathbf x_2 - \mathbf x_3}^2
\right)
\right]
\frac{1}{\norm{\mathbf r - \mathbf s}}
\\
&=
{\left(
\frac{\beta}{\alpha}
\right)}^3
\exp\left[-\frac{\alpha}{2} \left(\norm{\mathbf x_1 - \mathbf x_4}^2 + \norm{\mathbf x_2 - \mathbf x_3}^2\right)\right]
{\left(
\frac{2\alpha}{\pi}
\right)}^3
\int d \mathbf r d \mathbf s
\exp\left[
-\alpha \left(
2 \norm{\mathbf r}^2
+
2 \norm{\mathbf s - \frac{\mathbf x_2 + \mathbf x_3 - \mathbf x_1 - \mathbf x_4}{2}}^2
\right)
\right]
\frac{1}{\norm{\mathbf r - \mathbf s}}
\\
&=
{\left(
\frac{\beta}{\alpha}
\right)}^3
\exp\left[-\frac{\alpha}{2} \left(\norm{\mathbf x_1 - \mathbf x_4}^2 + \norm{\mathbf x_2 - \mathbf x_3}^2\right)\right]
\ham[u]{Coul}_{\alpha}\left(
\frac{\mathbf x_2 + \mathbf x_3 - \mathbf x_1 - \mathbf x_4}{2}
\right)
\\
&\leq \beta^3 \alpha^{-3} \sqrt{\frac{4 \alpha}{\pi}} 
\leq 2 
\beta^3 \alpha^{-5/2}
\leq
2
\beta^3
,
\label{eq:u-neg-bound-easy-case}
\end{align}
and so $u_{\max} \leq 2 \beta^3$.
To bound $\ham[u]{neg}_{\max}$, consider the above when at least one pair of the points $\mathbf x_1$ through $\mathbf x_4$ are at least $\Gamma$ apart.
If $\norm{\mathbf x_1 - \mathbf x_4} \geq \Gamma / 2$, then the integral is at most
\begin{align}
\beta^3
\alpha^{-3}
\exp\left[-\frac{\alpha}{2} \norm{\mathbf x_1 - \mathbf x_4}^2\right]
\ham[u]{Coul}_{\alpha}(0)
\leq
\beta^3
\alpha^{-3}
\exp\left(-\alpha \Gamma^2 / 8\right)
2 \sqrt{\alpha}
\leq
2 \beta^3
\exp\left(-\alpha \Gamma^2 / 8\right)
\end{align}
and similarly for $\norm{\mathbf x_2 - \mathbf x_3} \leq \Gamma / 2$.
If neither of these are the case then at least one of $\mathbf x_1, \mathbf x_4$ must be at least $\Gamma$ away from at least one of $\mathbf x_2, \mathbf x_3$.
Without loss of generality, suppose $\norm{\mathbf x_1 - \mathbf x_2} \geq \Gamma$.
That $\norm{\mathbf x_1 - \mathbf x_4} \leq \Gamma / 2$ implies
\begin{equation}
\norm{\frac{\mathbf x_1 + \mathbf x_4}{2} - \mathbf x_1} \leq \frac{\Gamma}{4}
\end{equation}
and similarly for $\mathbf x_2$ and $\mathbf x_3$.
Then
\begin{align}
\norm{\frac{\mathbf x_2 + \mathbf x_3 - \mathbf x_1 - \mathbf x_4}{2}}
=
\norm{
\left(
\underbrace{
\frac{\mathbf x_2 + \mathbf x_3}{2} - \mathbf x_2
}_{\leq \Gamma / 4}
\right)
-
\left(
\underbrace{
\frac{\mathbf x_1 + \mathbf x_4}{2} - \mathbf x_1
}_{\leq \Gamma / 4}
\right)
+
\left(
\underbrace{
\mathbf x_2 - \mathbf x_1
}_{\geq \Gamma}
\right)
}
\geq
\Gamma - \frac{\Gamma}{4} - \frac{\Gamma}{4} = \frac{\Gamma}{2}
\end{align}
and the potential integral is at most
\begin{align}
\beta^3 \ham[u]{Coul}_{\alpha}(\Gamma / 2)
=
\beta^3 \sqrt{\frac{4 \alpha}{\pi}} F_0 (\alpha \Gamma^2 / 4)
\leq
\beta^3 \sqrt{\frac{4 \alpha}{\pi}} \sqrt{\frac{\pi}{4}} \frac1{\sqrt{\alpha \Gamma^2 / 4}}
=
2 \beta^3 \frac{1}{\Gamma}
.
\end{align}
Together with~\cref{eq:u-neg-bound-easy-case}, this yields
\begin{align}
\ham[u]{neg}_{\max} \leq 
\max\left\{
2 \beta^3 \exp(-\alpha \Gamma^2 / 8)
,
2 \beta^3 \Gamma^{-1}
\right\}
=
2 \beta^3 \max\{\exp(-\alpha \Gamma^2 / 8), \Gamma^{-1}\}
\end{align}
For $\alpha \Gamma^2 \geq 64$ (a condition of the lemma),
\begin{align}
\frac{1}{\exp(-\alpha \Gamma^2 / 8)}
&=
\exp(\alpha \Gamma^2 / 8)
=
\sum_{k=0}^{\infty}
\frac{1}{k!}
{\left(
\frac{\alpha}{8} \Gamma^2
\right)}^k
\\
&
\leq
\frac{\alpha}{8} \Gamma^2
\leq
\frac{\alpha^2}{8} \Gamma^2
=
\sqrt{\frac{\alpha \Gamma^2}{64}}
\cdot
\Gamma
\geq
\Gamma
\end{align}
and so
\begin{equation}
\ham[u]{neg}_{\max} \leq 2 \beta^3 / \Gamma.
\end{equation}
\end{proof}

\bibliographystyle{plainurl}
\bibliography{complexity_of_es}

\end{document}